\documentstyle[12pt,psfig,epsfig,axodraw]{article}
\def\baselinestretch{1.0}
\advance\textheight by \topskip
\begin{document}
\tolerance=100000
\thispagestyle{empty}
\setcounter{page}{0}
\topmargin -0.1in
\headsep 30pt
\footskip 40pt
\oddsidemargin 12pt
\evensidemargin -16pt
\textheight 8.5in
\textwidth 6.5in
\parindent 20pt
 
\def\baselinestretch{1.5}
\newcommand{\newc}{\newcommand}
\def\preprint{{preprint}}
\def\Ord{\lower .7ex\hbox{$\;\stackrel{\textstyle <}{\sim}\;$}}
\def\OOrd{\lower .7ex\hbox{$\;\stackrel{\textstyle >}{\sim}\;$}}
\def\cO#1{{\cal{O}}\left(#1\right)}
\newc{\order}{{\cal O}}
\def\lag             {{\cal L}}
\def\Lag             {{\cal L}}
\def\lum             {{\cal L}}
\def\R               {{\cal R}}
\def\Rsq             {{\cal R}^{\sq}}
\def\Rst             {{\cal R}^{\st}}
\def\Rsb             {{\cal R}^{\sb}}
\def\M               {{\cal M}}
\def\Oas             {{\cal O}(\alpha_{s})}
\def\Vcal            {{\cal V}}
\def\Wcal            {{\cal W}}
\newc{\be}{\begin{equation}}
\newc{\ee}{\end{equation}}
\newc{\br}{\begin{eqnarray}}
\newc{\er}{\end{eqnarray}}
\newc{\ba}{\begin{array}}
\newc{\ea}{\end{array}}
\newc{\bi}{\begin{itemize}}
\newc{\ei}{\end{itemize}}
\newc{\bn}{\begin{enumerate}}
\newc{\en}{\end{enumerate}}
\newc{\bc}{\begin{center}}
\newc{\ec}{\end{center}}
\newc{\ul}{\underline}
\newc{\ol}{\overline}
\newc{\ra}{\rightarrow}
\newc{\lra}{\longrightarrow}
\newc{\wt}{\widetilde}
\newc{\til}{\tilde}
\def\kr              {^{\dagger}}
\newc{\wh}{\widehat}
\newc{\ti}{\times}
\newc{\Dir}{\kern -6.4pt\Big{/}}
\newc{\Dirin}{\kern -10.4pt\Big{/}\kern 4.4pt}
\newc{\DDir}{\kern -10.6pt\Big{/}}
\newc{\DGir}{\kern -6.0pt\Big{/}}
\newc{\sig}{\sigma}
\newc{\sigmalstop}{\sig_{\lstoppair}}
\newc{\Sig}{\Sigma}  
\newc{\del}{\delta}
\newc{\Del}{\Delta}
\newc{\lam}{\lambda}
\newc{\Lam}{\Lambda}
\newc{\gam}{\gamma}
\newc{\Gam}{\Gamma}
\newc{\eps}{\epsilon}
\newc{\Eps}{\Epsilon}
\newc{\kap}{\kappa}
\newc{\Kap}{\Kappa}
\newc{\modulus}[1]{\left| #1 \right|}
\newc{\eq}[1]{(\ref{eq:#1})}
\newc{\eqs}[2]{(\ref{eq:#1},\ref{eq:#2})}
\newc{\etal}{{\it et al.}\ }
\newc{\ibid}{{\it ibid}.}
\newc{\ibidem}{{\it ibidem}.}
\newc{\eg}{{\it e.g.}\ }
\newc{\ie}{{\it i.e.}\ }
\def \viz{\emph{viz.}}
\def \etc{\emph{etc. }}
\newc{\nonum}{\nonumber}
\newc{\lab}[1]{\label{eq:#1}}
\newc{\dpr}[2]{({#1}\cdot{#2})}
\newc{\lt}{\stackrel{<}}
\newc{\gt}{\stackrel{>}}
\newc{\lsimeq}{\stackrel{<}{\sim}}
\newc{\gsimeq}{\stackrel{>}{\sim}}
\def\lsim{\buildrel{\scriptscriptstyle <}\over{\scriptscriptstyle\sim}}
\def\gsim{\buildrel{\scriptscriptstyle >}\over{\scriptscriptstyle\sim}}
\def\lapp{\mathrel{\rlap{\raise.5ex\hbox{$<$}}
                    {\lower.5ex\hbox{$\sim$}}}}
\def\gapp{\mathrel{\rlap{\raise.5ex\hbox{$>$}}
                    {\lower.5ex\hbox{$\sim$}}}}
\newc{\half}{\frac{1}{2}}
\newcommand {\nnc}        {{\overline{\mathrm N}_{95}}}
\newcommand {\dm}         {\Delta m}
\newcommand {\dM}         {\Delta M}
\def\bra{\langle}
\def\ket{\rangle}
\def\cO#1{{\cal{O}}\left(#1\right)}
\def \DM{{\Delta{m}}}
\newc{\bQ}{\ol{Q}}
\newc{\dota}{\dot{\alpha }}
\newc{\dotb}{\dot{\beta }}
\newc{\dotd}{\dot{\delta }}
\newc{\nindnt}{\noindent}

\newcommand{\medf}[2] {{\footnotesize{\frac{#1}{#2}} }}
\newcommand{\smaf}[2] {{\textstyle \frac{#1}{#2} }}
\def\onesq            {{\textstyle \frac{1}{\sqrt{2}} }}
\def\onehf            {{\textstyle \frac{1}{2} }}
\def\oneth            {{\textstyle \frac{1}{3} }}
\def\twoth            {{\textstyle \frac{2}{3} }}
\def\onefo            {{\textstyle \frac{1}{4} }}
\def\forth            {{\textstyle \frac{4}{3} }}

\newc{\matth}{\mathsurround=0pt}
\def\ML{\ifmmode{{\mathaccent"7E M}_L}
             \else{${\mathaccent"7E M}_L$}\fi}
\def\MR{\ifmmode{{\mathaccent"7E M}_R}
             \else{${\mathaccent"7E M}_R$}\fi}
\newcommand{\s}{\\ \vspace*{-3mm} }

\def \ud { {1 \over 2} }
\def \ut { {1 \over 3} }
\def \td { {3 \over 2} }
\newc{\mr}{\mathrm}
\def\dh {\partial }
\def \cs { cross-section }
\def \css { cross-sections }
\def \cm { centre of mass }
\def \cms { centre of mass energy }
\def \cc { coupling constant }
\def \ccs {coupling constants }
\def \gc {gauge coupling }
\def \gcc {gauge coupling constant }
\def \gccs {gauge coupling constants }
\def \yc {Yukawa coupling }
\def \ycc {Yukawa coupling constant }
\def \pp {{parameter }}
\def \pps {{parameters }} 
\def \ps {parameter space }
\def \pss {parameter spaces }
\def \vv {vice versa }

\newc{\siminf}{\mbox{$_{\sim}$ {\small {\hspace{-1.em}{$<$}}}    }}
\newc{\simsup}{\mbox{$_{\sim}$ {\small {\hspace{-1.em}{$>$}}}    }}


\newc {\Zboson}{{\mathrm Z}^{0}}
\newc{\thetaw}{\theta_W}
\newc{\mbot}{{m_b}}
\newc{\mtop}{{m_t}}
\newc{\sm}{${\cal {SM}}$}
\newc{\as}{\alpha_s}
\newc{\aem}{\alpha_{em}}
\def \PI{{\pi^{\pm}}}
\newc{\ppbar}{\mbox{$p\ol{p}$}}
\newc{\bbbar}{\mbox{$b\ol{b}$}}
\newc{\ccbar}{\mbox{$c\ol{c}$}}
\newc{\ttbar}{\mbox{$t\ol{t}$}}
\newc{\eebar}{\mbox{$e\ol{e}$}}
\newc{\zzero}{\mbox{$Z^0$}}
\def \gamz{\Gam_Z}
\newc{\wplus}{\mbox{$W^+$}}
\newc{\wminus}{\mbox{$W^-$}}
\newc{\ellp}{\ell^+}
\newc{\ellm}{\ell^-}
\newc{\elp}{\mbox{$e^+$}}
\newc{\elm}{\mbox{$e^-$}}
\newc{\elpm}{\mbox{$e^{\pm}$}}
\newc{\qbar}     {\mbox{$\ol{q}$}}
\def \ewgroup{SU(2)_L \otimes U(1)_Y}
\def \smgroup{SU(3)_C \otimes SU(2)_L \otimes U(1)_Y}
\def \smcolorem{SU(3)_C \otimes U(1)_{em}}

\def \SSM  {Supersymmetric Standard Model}
\def \poincare{Poincare$\acute{e}$}
\def \superspace{\emph{superspace}}
\def \sfs{\emph{superfields}}
\def \superpot{\emph{superpotential}}
\def \csf{\emph{chiral superfield}}
\def \csfs{\emph{chiral superfields}}
\def \vsf{\emph{vector superfield }}
\def \vsfs{\emph{vector superfields}}
\newc{\Ebar}{{\bar E}}
\newc{\Dbar}{{\bar D}}
\newc{\Ubar}{{\bar U}}
\newc{\susy}{{{SUSY}}}
\newc{\msusy}{{{M_{SUSY}}}}

\def\photino{\ifmmode{\mathaccent"7E \gam}\else{$\mathaccent"7E \gam$}\fi}
\def\taugluino{\ifmmode{\tau_{\mathaccent"7E g}}
             \else{$\tau_{\mathaccent"7E g}$}\fi}
\def\mphotino{\ifmmode{m_{\mathaccent"7E \gam}}
             \else{$m_{\mathaccent"7E \gam}$}\fi}
\newc{\gl}   {\mbox{$\wt{g}$}}
\newc{\mgl}  {\mbox{$m_{\gl}$}}
\def \charginopm{{\wt\chi}^{\pm}}
\def \mcharginopm{m_{\charginopm}}
\def \mchpmmin {\mcharginopm^{min}}
\def \chonep {{\wt\chi_1^+}}
\def \chonem {{\wt\chi_1^-}}
\def \chplus {{\wt\chi^+}}
\def \chminus {{\wt\chi^-}}
\def \chonip{{\wt\chi_i}^{+}}
\def \chonim{{\wt\chi_i}^{-}}
\def \chonipm{{\wt\chi_i}^{\pm}}
\def \chonjp{{\wt\chi_j}^{+}}
\def \chonjm{{\wt\chi_j}^{-}}
\def \chonjpm{{\wt\chi_j}^{\pm}}
\def \chonepm{{\wt\chi_1}^{\pm}}
\def \chonemp{{\wt\chi_1}^{\mp}}
\def \mchonepm{m_{\chonepm}}
\def \mchonemp{m_{\chonemp}}
\def \chtwopm{{\wt\chi_2}^{\pm}}
\def \mchtwopm{m_{\chtwopm}}
\newc{\dmchi}{\Delta m_{\wt\chi}}


\def \vlsp{\emph{VLSP}}
\def \lspi{\wt\chi_i^0}
\def \mlspi{m_{\lspi}}
\def \lspj{\wt\chi_j^0}
\def \mlspj{m_{\lspj}}
\def \lspone{\wt\chi_1^0}
\def \mlspone{m_{\lspone}}
\def \lsptwo{\wt\chi_2^0}
\def \mlsptwo{m_{\lsptwo}}
\def \lspthree{\wt\chi_3^0}
\def \mlspthree{m_{\lspthree}}
\def \lspfour{\wt\chi_4^0}
\def \mlspfour{m_{\lspfour}}


\newc{\sele}{\wt{\mathrm e}}
\newc{\sell}{\wt{\ell}}
\def \msell{m_{\sell}}
\def \slepone{\wt\ell_1}
\def \mslepone{m_{\slepone}}
\def \smuone{\wt\mu_1}
\def \msmuone{m_{\smuone}}
\def \stauone{\wt\tau_1}
\def \mstauone{m_{\stauone}}
\def \snu{\wt{\nu}}
\def \msnu{m_{\snu}}
\def \msnumu{m_{\snu_{\mu}}}
\def \barsnu{\wt{\bar{\nu}}}
\def \barsnul{\barsnu_{\ell}}
\def \snul{\snu_{\ell}}
\def \mbarsnu{m_{\barsnu}}
\newc{\snue}     {\mbox{$ \wt{\nu_e}                         $}}
\newc{\smu}{\wt{\mu}}
\newc{\stau}{\wt{\tau}}
\newc {\nuL} {\wt{\nu}_L}
\newc {\nuR} {\wt{\nu}_R}
\newc {\snub} {\bar{\wt{\nu}}}
\newc {\eL} {\wt{e}_L}
\newc {\eR} {\wt{e}_R}
\def \slepl{\wt{l}_L}
\def \mslepl{m_{\slepl}}
\def \slepr{\wt{l}_R}
\def \mslepr{m_{\slepr}}
\def \stau{\wt\tau}
\def \mstau{m_{\stau}}
\def \slepton{\wt\ell}
\def \mslepton{m_{\slepton}}
\def \mlhiggs{m_{h^0}}

\def \xr{X_{r}}

\def \sfer{\wt{f}}
\def \msfer{m_{\sfer}}
\def \sq{\wt{q}}
\def \msq{m_{\sq}}
\def \msquleft{m_{\tilde{u_L}}}
\def \msqurht{m_{\tilde{u_R}}}
\def \sql{\wt{q}_L}
\def \msql{m_{\sql}}
\def \sqr{\wt{q}_R}
\def \msqr{m_{\sqr}}
\newc{\msqot}  {\mbox{$m_(\sq_{1,2} )$}}
\newc{\sqbar}    {\mbox{$\bar{\wt{q}}$}}
\newc{\ssb}      {\mbox{$\squark\ol{\squark}$}}
\newc {\qL} {\wt{q}_L}
\newc {\qR} {\wt{q}_R}
\newc {\uL} {\wt{u}_L}
\newc {\uR} {\wt{u}_R}
\def \ul{\wt{u}_L}
\def \mul{m_{\ul}}
\newc {\dL} {\wt{d}_L}
\newc {\dR} {\wt{d}_R}
\newc {\cL} {\wt{c}_L}
\newc {\cR} {\wt{c}_R}
\newc {\sL} {\wt{s}_L}
\newc {\sR} {\wt{s}_R}
\newc {\tL} {\wt{t}_L}
\newc {\tR} {\wt{t}_R}
\newc {\stb} {\ol{\wt{t}}_1}
\newc {\sbot} {\wt{b}_1}
\newc {\msbot} {m_{\sbot}}
\newc {\sbotb} {\ol{\wt{b}}_1}
\newc {\bL} {\wt{b}_L}
\newc {\bR} {\wt{b}_R}
\def \mul{m_{\wt{u}_L}}
\def \mur{m_{\wt{u}_R}}
\def \mdl{m_{\wt{d}_L}}
\def \mdr{m_{\wt{d}_R}}
\def \mcl{m_{\wt{c}_L}}
\def \charml{\wt{c}_L}
\def \mcr{m_{\wt{c}_R}}
\newc{\csquark}  {\mbox{$\wt{c}$}}
\newc{\csquarkl} {\mbox{$\wt{c}_L$}}
\newc{\mcsl}     {\mbox{$m(\csquarkl)$}}
\def \msl{m_{\wt{s}_L}}
\def \msr{m_{\wt{s}_R}}
\def \mbl{m_{\wt{b}_L}}
\def \mbr{m_{\wt{b}_R}}
\def \mtl{m_{\wt{t}_L}}
\def \mtr{m_{\wt{t}_R}}
\def \st{\wt{t}}
\def \mst{m_{\st}}
\newc {\stopl}         {\wt{\mathrm{t}}_{\mathrm L}}
\newc {\stopr}         {\wt{\mathrm{t}}_{\mathrm R}}
\newc {\stoppair}      {\wt{\mathrm{t}}_{1}
\bar{\wt{\mathrm{t}}}_{1}}
\def \lstop{\wt{t}_{1}}
\def \lstopbar{\lstop^*}
\def \hstop{\wt{t}_{2}}
\def \hstopbar{\hstop^*}
\def \mlstop{m_{\lstop}}
\def \mhstop{m_{\hstop}}
\def \lstoppair{\lstop\lstop^*}
\def \hstoppair{\hstop\hstop^*}
\newc{\tsquark}  {\mbox{$\wt{t}$}}
\newc{\ttb}      {\mbox{$\tsquark\ol{\tsquark}$}}
\newc{\ttbone}   {\mbox{$\tsquark_1\ol{\tsquark}_1$}}
\def \tsq {top squark }
\def \tsqs {top squarks }
\def \tsql {ligtest top squark }
\def \tsqh {heaviest top squark }
\newc{\mix}{\theta_{\wt t}}
\newc{\cost}{\cos{\theta_{\wt t}}}
\newc{\sint}{\sin{\theta_{\wt t}}}
\newc{\costloop}{\cos{\theta_{\wt t_{loop}}}}
\def \lsbot{\wt{b}_{1}}
\def \lsbotbar{\lsbot^*}
\def \hsbot{\wt{b}_{2}}
\def \hsbotbar{\hsbot^*}
\def \mlsbot{m_{\lsbot}}
\def \mhsbot{m_{\hsbot}}
\def \lsbotpair{\lsbot\lsbot^*}
\def \hsbotpair{\hsbot\hsbot^*}
\newc{\mixsbot}{\theta_{\wt b}}

\def \mhone{m_{h_1}}
\def \hup{{H_u}}
\def \hdn{{H_d}}
\newc{\tb}{\tan\beta}
\newc{\cb}{\cot\beta}
\newc{\vev}[1]{{\left\langle #1\right\rangle}}

\def \abot{A_{b}}
\def \atop{A_{t}}
\def \atau{A_{\tau}}
\newc{\mhalf}{m_{1/2}}
\newc{\mzero} {\mbox{$m_0$}}
\newc{\azero} {\mbox{$A_0$}}

\newc{\lb}{\lam}
\newc{\lbp}{\lam^{\prime}}
\newc{\lbpp}{\lam^{\prime\prime}}
\newc{\rpv}{{\not \!\! R_p}}
\newc{\rpvm}{{\not  R_p}}
\newc{\rp}{R_{p}}
\newc{\rpmssm}{{RPC MSSM}}
\newc{\rpvmssm}{{RPV MSSM}}


\newc{\sbyb}{S/$\sqrt B$}
\newc{\pelp}{\mbox{$e^+$}}
\newc{\pelm}{\mbox{$e^-$}}
\newc{\pelpm}{\mbox{$e^{\pm}$}}
\newc{\epem}{\mbox{$e^+e^-$}}
\newc{\lplm}{\mbox{$\ell^+\ell^-$}}
\def \branch{\emph{BR}}
\def \branche{\branch(\lstop\ra be^{+}\nu_e \lspone)\ti \branch(\lstop^{*}\ra \bar{b}q\bar{q^{\prime}}\lspone)}
\def \branchmu{\branch(\lstop\ra b\mu^{+}\nu_{\mu} \lspone)\ti \branch(\lstop^{*}\ra \bar{b}q\bar{q^{\prime}}\lspone)}
\def\Ecm{\ifmmode{E_{\mathrm{cm}}}\else{$E_{\mathrm{cm}}$}\fi}
\newc{\rts}{\sqrt{s}}
\newc{\rtshat}{\sqrt{\hat s}}
\newc{\gev}{\,GeV}
\newc{\mev}{~{\rm MeV}}
\newc{\tev}  {\mbox{$\;{\rm TeV}$}}
\newc{\gevc} {\mbox{$\;{\rm GeV}/c$}}
\newc{\gevcc}{\mbox{$\;{\rm GeV}/c^2$}}
\newc{\intlum}{\mbox{${ \int {\cal L} \; dt}$}}
\newc{\call}{{\cal L}}
\def \met  {\mbox{${E\!\!\!\!/_T}$}}
\def \cpv  {\mbox{${CP\!\!\!\!/}$}}
\newc{\ptmiss}{/ \hskip-7pt p_T}
\def \eslash{\not \! E}
\def \etslash{\not \! E_T }
\def \ptslash{\not \! p_T }
\newc{\PT}{\mbox{$p_T$}}
\newc{\ET}{\mbox{$E_T$}}
\newc{\dedx}{\mbox{${\rm d}E/{\rm d}x$}}
\newc{\ifb}{\mbox{${\rm fb}^{-1}$}}
\newc{\ipb}{\mbox{${\rm pb}^{-1}$}}
\newc{\pb}{~{\rm pb}}
\newc{\fb}{~{\rm fb}}
\newc{\ycut}{y_{\mathrm{cut}}}
\newc{\chis}{\mbox{$\chi^{2}$}}
\def \hadron{\emph{hadron}}
\def \nlc{\emph{NLC }}
\def \lhc{\emph{LHC }}
\def \cdf{\emph{CDF }}
\def\dzero{\emptyset}
\def \tevatron{\emph{Tevatron }}
\def \lep{\emph{LEP }}
\def \jets{\emph{jets }}
\def \jet(s){\emph{jet(s) }}

\def\Crs{stroke [] 0 setdash exch hpt sub exch vpt add hpt2 vpt2 neg V currentpoint stroke 
hpt2 neg 0 R hpt2 vpt2 V stroke}
\def\loopdk{\lstop \ra c \lspone}
\def\brloopdk{\branch(\loopdk)}
\def\fourdk{\lstop \ra b \lspone  f \bar f'}
\def\brfourdk{\branch(\fourdk)}
\def\fourdklep{\lstop \ra b \lspone  \ell \nu_{\ell}}
\def\fourdkhad{\lstop \ra b \lspone  q \bar q'}
\def\brfourdklep{\branch(\fourdklep)}
\def\brfourdkhad{\branch(\fourdkhad)}
\def\twodk{\lstop \ra b \chonep}
\def\brtwodk{\branch(\twodk)}
\def\threedkslep{\lstop \ra b \wt{\ell} \nu_{\ell}}
\def\brthreedkslep{\branch(\threedkslep)}
\def\threedksnu{\lstop \ra b \wt{\nu_{\ell}} \ell }
\def\brthreedksnu{\branch(\threedksnu) }
\def\threedklsp{\lstop \ra b W \lspone }
\def\brthreedklsp{\\branch(\threedklsp) }
\def\topdk{t \ra \lstop \lspone}
\def\rpvdk{\lstop \ra e^+ d}
\def\brrpvdk{\branch(\rpvdk)}
\def\fonec{f_{11c}} 
\newc{\mpl}{M_{\rm Pl}}
\newc{\mgut}{M_{GUT}}
\newc{\mw}{M_{W}}
\newc{\mweak}{M_{weak}}
\newc{\mz}{M_{Z}}

\newc{\OPALColl}   {OPAL Collaboration}
\newc{\ALEPHColl}  {ALEPH Collaboration}
\newc{\DELPHIColl} {DELPHI Collaboration}
\newc{\XLColl}     {L3 Collaboration}
\newc{\JADEColl}   {JADE Collaboration}
\newc{\CDFColl}    {CDF Collaboration}
\newc{\DXColl}     {D0 Collaboration}
\newc{\HXColl}     {H1 Collaboration}
\newc{\ZEUSColl}   {ZEUS Collaboration}
\newc{\LEPColl}    {LEP Collaboration}
\newc{\ATLASColl}  {ATLAS Collaboration}
\newc{\CMSColl}    {CMS Collaboration}
\newc{\UAColl}    {UA Collaboration}
\newc{\KAMLANDColl}{KamLAND Collaboration}
\newc{\IMBColl}    {IMB Collaboration}
\newc{\KAMIOColl}  {Kamiokande Collaboration}
\newc{\SKAMIOColl} {Super-Kamiokande Collaboration}
\newc{\SUDANTColl} {Soudan-2 Collaboration}
\newc{\MACROColl}  {MACRO Collaboration}
\newc{\GALLEXColl} {GALLEX Collaboration}
\newc{\GNOColl}    {GNO Collaboration}
\newc{\SAGEColl}  {SAGE Collaboration}
\newc{\SNOColl}  {SNO Collaboration}
\newc{\CHOOZColl}  {CHOOZ Collaboration}
\newc{\PDGColl}  {Particle Data Group Collaboration}

\def\issue(#1,#2,#3){{\bf #1} (#3) #2 } 
\def\ASTR(#1,#2,#3){Astropart.\ Phys. \issue(#1,#2,#3)}
\def\AJ(#1,#2,#3){Astrophysical.\ Jour. \issue(#1,#2,#3)}
\def\AJS(#1,#2,#3){Astrophys.\ J.\ Suppl. \issue(#1,#2,#3)}
\def\APP(#1,#2,#3){Acta.\ Phys.\ Pol. \issue(#1,#2,#3)}
\def\JCAP(#1,#2,#3){Journal\ XX. \issue(#1,#2,#3)} 
\def\SC(#1,#2,#3){Science \issue(#1,#2,#3)}
\def\PRD(#1,#2,#3){Phys.\ Rev.\ D \issue(#1,#2,#3)}
\def\PR(#1,#2,#3){Phys.\ Rev.\ \issue(#1,#2,#3)} 
\def\PRC(#1,#2,#3){Phys.\ Rev.\ C \issue(#1,#2,#3)}
\def\NPB(#1,#2,#3){Nucl.\ Phys.\ B \issue(#1,#2,#3)}
\def\NPPS(#1,#2,#3){Nucl.\ Phys.\ Proc. \ Suppl \issue(#1,#2,#3)}
\def\NJP(#1,#2,#3){New.\ J.\ Phys. \issue(#1,#2,#3)}
\def\JP(#1,#2,#3){J.\ Phys.\issue(#1,#2,#3)}
\def\PL(#1,#2,#3){Phys.\ Lett. \issue(#1,#2,#3)}
\def\PLB(#1,#2,#3){Phys.\ Lett.\ B  \issue(#1,#2,#3)}
\def\ZP(#1,#2,#3){Z.\ Phys. \issue(#1,#2,#3)}
\def\ZPC(#1,#2,#3){Z.\ Phys.\ C  \issue(#1,#2,#3)}
\def\PREP(#1,#2,#3){Phys.\ Rep. \issue(#1,#2,#3)}
\def\PRL(#1,#2,#3){Phys.\ Rev.\ Lett. \issue(#1,#2,#3)}
\def\MPL(#1,#2,#3){Mod.\ Phys.\ Lett. \issue(#1,#2,#3)}
\def\RMP(#1,#2,#3){Rev.\ Mod.\ Phys. \issue(#1,#2,#3)}
\def\SJNP(#1,#2,#3){Sov.\ J.\ Nucl.\ Phys. \issue(#1,#2,#3)}
\def\CPC(#1,#2,#3){Comp.\ Phys.\ Comm. \issue(#1,#2,#3)}
\def\IJMPA(#1,#2,#3){Int.\ J.\ Mod. \ Phys.\ A \issue(#1,#2,#3)}
\def\MPLA(#1,#2,#3){Mod.\ Phys.\ Lett.\ A \issue(#1,#2,#3)}
\def\PTP(#1,#2,#3){Prog.\ Theor.\ Phys. \issue(#1,#2,#3)}
\def\RMP(#1,#2,#3){Rev.\ Mod.\ Phys. \issue(#1,#2,#3)}
\def\NIMA(#1,#2,#3){Nucl.\ Instrum.\ Methods \ A \issue(#1,#2,#3)}
\def\JHEP(#1,#2,#3){J.\ High\ Energy\ Phys. \issue(#1,#2,#3)}
\def\EPJC(#1,#2,#3){Eur.\ Phys.\ J.\ C \issue(#1,#2,#3)}
\def\RPP (#1,#2,#3){Rept.\ Prog.\ Phys. \issue(#1,#2,#3)}
\def\PPNP(#1,#2,#3){ Prog.\ Part.\ Nucl.\ Phys. \issue(#1,#2,#3)}
\newc{\PRDR}[3]{{Phys. Rev. D} {\bf #1}, Rapid  Communications, #2 (#3)}

\vspace*{\fill}
\vspace{-0.5in}
\begin{flushright}
{\tt hep-ph/0512011}\\
{\tt JU-PHYSICS/12/05}
\end{flushright}
\begin{center}
{\Large \bf
Four body decay of the lighter top-squark constrained by the Lighter  
CP-even Higgs boson mass bound  
}\\[1.00
cm]
{\large Siba Prasad Das$^{}$\footnote{\it spdas@juphys.ernet.in}
}\\[0.3 cm]
{\it  Department of Physics, Jadavpur University,
Kolkata- 700 032, India}\\[0.3cm]
\end{center}
\vspace{.2cm}

\begin{abstract}
{\noindent \normalsize
We reinvestigated the parameter space allowing  a large BR of 
the 4-body decay of the lighter top-squark ($\lstop$) accessible 
at Tevatron Run-II by imposing  
the lighter CP-even Higgs boson mass ($\mlhiggs$) bound from LEP. 
Important constraints were obtained  
in mSUGRA as well as in the unconstrained supersymmetric models. Our results 
show that the prospect of searching the lighter top-squark via the 
4-body decay  mode, in particular the  $\ell + n-jets + \met$ signal, is not 
promising 
in mSUGRA due to the above  bound on $\mlhiggs$. The existing 
bounds on $\mlstop$ from Tevatron Run-I and LEP assuming 100\% BR of the 
loop decay of $\lstop$ are, therefore, 
valid to a good approximation. We also find that large BRs of the above  
4-body decay are allowed in the unconstrained model 
over significant regions of parameter spaces and the possibility that
this decay mode is the main discovery channel at Tevatron Run-II is open. 
We have briefly reviewed
the theoretical uncertainties in the calculation of $\mlhiggs$ and
their consequences for the constraints obtained by us.
We have commented 
upon, with illustrative examples, how the above parameter space 
is affected if future experiments push the Higgs boson 
mass bound upward. }

\end{abstract}
\vskip1.0cm
\noindent
\vspace*{\fill}
\newpage

\section{Introduction} 
\label{intro4}

The mass of the lighter top-squark ($\lstop$), superpartner of the 
top-quark in the standard model (SM)  may turn out to be below the top-quark mass. 
This happens in a wide region 
of the Supersymmetry (SUSY) parameter space both in the 
minimal Supergravity (mSUGRA) model and in the  
Minimal Supersymmetric Standard Model (MSSM). It could even be the 
next-to-lightest supersymmetric particle (NLSP), the lightest neutralino 
($\lspone$) being the lightest supersymmetric particle (LSP) by 
assumption. Being the lightest among the strongly interacting sparticles 
with a relatively large cross-section compared to other sparticles, the lighter 
top-squark may show up in direct searches at Tevatron Run-II. In our  
convention,  $\lstop = \stopl \cost + \stopr \sint $ and  
$\hstop = $-$\stopl \sint + \stopr \cost $, where $\mix$ is the 
mixing angle and  $\hstop$ is the heavier top-squark. The top-quark mass ($m_t$)  
appears in the off-diagonal elements of the top-squark mass matrix. So, 
the physical top-squark masses ($\mlstop$, $\mhstop$) and the mixing 
angle depend on the input value of the top-quark mass.

In the lighter top-squark-NLSP ($\lstop$-NLSP) scenario   
($\ie$, $\mlspone < \mlstop < $ the mass of any other sparticle), $\lstop$ has few allowed 
decay modes in the R-parity conserving (RPC) SUSY model. They are the 
Flavor Changing Neutral Current (FCNC) induced loop decay, $\loopdk$ 
\cite{hikasa}, the 4-body decay, $\fourdk$, 
where $f \bar f'=u,d,c,s,\ell,\nu_{\ell}$ \cite{boehm}, the  
tree-level two body decay, $\lstop \ra t \lspone$ and 
the three body decay, $\lstop \ra b W \lspone$. If $\mlstop \gsim m_{t} + 
\mlspone$, the 
production cross-section of $\lstop$ is rather low for the Tevatron 
experiments. The $\lstop \ra b W \lspone$ mode may be 
suppressed if the coupling $W \lspone \chonepm$ is small or 
$\mlstop < \mw + \mlspone + \mbot$. This
may happen if $\lspone$ is bino like as in the mSUGRA model. 
However, in this report we have considered only those 
parameter spaces where the last two decay modes are 
either  kinematically  or dynamically suppressed.

So far most of the $\lstop$-NLSP searches at Tevatron Run-I and 
at the Large Electron Positron (LEP) collider have been performed 
assuming 100\% BR of the loop decay\footnote{Recently
the D$\dzero$  collaboration \cite{dzerostop} and the ALEPH
collaboration \cite{alephstoptopsqk} looked for the $\fourdk$ channel with special assumptions about its BR.}, 
leading to the signal: jets + missing energy ($\met$). The most stringent constraint comes from the Tevatron Run-I experiments \cite{tevstop}.
However, the  above assumption becomes unrealistic, if $\lstop$ has 
other decay modes with comparable widths. This possibility exists 
in a large region of the SUSY parameter spaces in both the models 
mentioned above \cite{boehm,me1}. The 4-body decay and the loop decay in RPC models 
may have  decay widths of the same order of magnitude in a wide 
region of the parameter space \cite{boehm}. 

An interesting feature of all supersymmetric models is the prediction of the 
existence of at least one light Higgs boson \cite{predhiggs,higgs135}. So far the 
LEP and Tevatron Run-I experiments found no significant evidence 
in favor of a light Higgs boson. The direct searches from the four 
LEP experiments concluded that the lighter Higgs boson must be heavier 
than 114.4 GeV at 95\% C.L. \cite{lephiggsjuly05}. Generally this 
bound depends on the value of sin$^2(\beta -\alpha)$, 
where $\beta$ and $\alpha$  are the 
ratio of the vacuum expectation values (VEVs) of the two Higgs doublets and 
the mixing angle in the CP-even Higgs boson mass matrix respectively. However, 
the Higgs boson mass bound is 114.4 GeV when sin$^2(\beta -\alpha)$ $\simeq$ 1. 
We have found that this factor is indeed $\simeq$ 1 for 
the parameter space interesting for us. This 
Higgs boson mass bound is directly applicable in SUSY when the CP-odd Higgs 
boson ($M_A$) mass is very large compared to Z-boson mass ($M_Z$), 
$\ie$, $M_A >> M_Z$, commonly known as the decoupling limit.

In SUSY the mass of the light Higgs boson at tree-level is given by,
\begin{eqnarray}
\mlhiggs^2&=&{1\over 2}\left [M_Z^2+M_A^2 - \sqrt{(M_Z^2+M_A^2)^2
-4M_Z^2M_A^2\cos ^22\beta}\right ]. \label{higgsmasscpeven} 
\end{eqnarray}

Eqn.\ref{higgsmasscpeven} shows that $\mlhiggs \lsim \mz$ at the tree-level.
But radiative corrections involving top
quark and squarks ($\wt t$) in the loop indicate that $\mlhiggs^2$ grows like 
$G_F N_c C m_t^4$ where 
$G_F$ is the Fermi coupling constant, $N_c$ is the color 
factor and C is a model dependent loop factor. 
 It is clear that the relatively large experimental error in 
 top mass may lead to a sizable uncertainty in the
Higgs boson mass prediction. The 
loop corrections from bottom quark and squarks ($\wt b $)
are also sizable for large values of $\tb$ and $\mu$ \cite{hepph0411114}, where  
$\mu$ is the Higgs mass parameter in the superpotential. We shall discuss the 
Higgs boson mass as a function of the SUSY model parameters in section \ref{numerical}.

Theoretically the mass of the lighter CP-even Higgs boson is bounded 
from  above: $\mlhiggs \lsim 135$ GeV. This bound has been obtained 
by including radiative corrections up to two-loop 
level \cite{higgs135,mhiggsletter,mhiggslong,mhiggsRG1a,mhiggsRG1b,mhiggsRG2}. 
So, $\mlhiggs$ is expected to lie in the range 114.4 GeV 
$\lsim \mlhiggs \lsim$ 135 GeV. Due to the yet unknown higher-order 
corrections the theoretical error on the lighter CP-even Higgs boson mass 
consists of several pieces \cite{sven1,sven2,sven3,sven4,djouadi}. They are the
 momentum-independent two-loop corrections, the 
momentum-dependent two-loop corrections, the  higher loop corrections 
from t/$\wt t $ sector etc. 
 Taking into account of all shorts of uncertainties 
the intrinsic error has been estimated to be $\delta \mlhiggs \approx$ 3 
GeV
\footnote {Inclusion of  the full two-loop corrections as well as the 
leading higher loop corrections is expected to yield a reduced uncertainty 
 $\delta \mlhiggs \lsimeq$ 0.5 GeV \cite{sven1,sven3} }. However 
the magnitude  of this uncertainty crucially depends on the 
SUSY parameter space.  In view of the above theoretical errors the 
viability of a parameter space will be judged in this work by using the 
following two constraints on the calculated Higgs 
mass: $(i)$ $\mlhiggs \gsim$ 111.4 
GeV (the weaker bound) $(ii)$ $\mlhiggs \gsim$ 114.4 GeV 
(the experimental bound). The first bound obviously leads to more 
conservative results as will be shown in section\ref{numerical}.

We now make a few comments on the $\lstop$ decay widths considered 
in this paper. 
In the approximation of neglecting the charm quark mass, 
$\tilde c_R$ does not mix with top-squarks . The mixing
of $\tilde c_L$ with the top-squarks mass eigenstates result in a redefined  
lighter top-squark given 
approximately by  $\tilde t_1$=$\tilde t_1$ +  $\epsilon \tilde c_L$, where 
\begin{eqnarray}
\epsilon = \frac{\Delta_L \cost-\Delta_R \sint}
 {m^2_{\tilde t_1}-m^2_{\tilde c_L}}.\qquad
\label{loopepsilon}
\end{eqnarray}

The parameters  $\Delta_{L,R}$ are given by
\begin{eqnarray}
\Delta_L &=& \frac{\alpha}{4 \pi s_W^2} \log \left(\frac{M_{\rm GUT}^2}
{M_W^2} \right)  \, \frac{V^*_{tb} V_{cb} \, m_b^2}{2M_W^2 \cos^2 \beta}
\bigg(m_{\tilde{c}_L}^2+m_{\tilde{b}_R}^2 + m_{H_d}^2 +A_b^2\bigg)
\label{loopdkL}
\end{eqnarray}
\begin{eqnarray}
\Delta_R &=&  \frac{\alpha}{4 \pi s_W^2} \log \left(\frac{M_{\rm GUT}^2}
{M_W^2} \right) \, \frac{V^*_{tb} V_{cb} \, m_b^2}{2M_W^2 \cos^2 \beta}m_t A_b,  
\label{loopdkLR}
\end{eqnarray}
where  $m_{\tilde{c}_L}$, $m_{\tilde{b}_R}$, $m_{H_d}$ and $A_b$ are the 
left-handed scharm, right-handed sbottom, down-type Higgs scalar 
and trilinear sbottom mass parameters respectively. $M_{\rm GUT}$ is the  
Grand Unification theory (GUT) scale in SUSY model. 


Assuming  proper electro-weak symmetry breaking (EWSB),
the Higgs scalar mass squared parameter (in Eqn.\ref{loopdkL})
can be written as  
\be
m_{H_d}^2= M_A^2 \sin^2\beta - \cos 2\beta M_W^2 -\mu^2 .
\label{ewsbmssm}
\ee 

The loop decay width can then be expressed as \cite{hikasa}:

\be
\Gamma( \tilde{t}_1 \ra c \chi_1^0) = \frac{\alpha}{4} m_{\tilde{t}_1} \
\left(1- \frac{m_{\chi_1^0}^2} {m_{\tilde{t}_1}^2 } \right)^2 |f_{L1}|^2 \
|\epsilon|^2
\label{loopdkwidth}
\ee

where $f_{L1}$ is given by
\be
f_{L1}  =  \sqrt{2}\left[ \frac{2}{3} (c_W N_{11} +s_W N_{12})
              + \left(\frac{1}{2} - \frac{2}{3}\, s_W^2 \right)
                 \frac{-s_W N_{11} +c_W N_{12}}{c_W s_W}\right]
\label{loopdkfl}
\ee
Here $N_{11}$ and $N_{12}$ denote the elements of 
the neutralino mixing matrix
 projecting the LSP on to the photino ($j=1$) and zino ($j=2$) states.
After renormalization one obtains the residual logarithmic terms in
Eqns.\ref{loopdkL} and \ref{loopdkLR} which  lead to a large contribution in the
amplitude. Adding various diagrams contributing to  the loop,
a divergent term  is obtained in general. In principle this
divergence must be cancelled  by adding appropriate counterterms.  
The dominant effects of renormalization are expected to come 
from the logarithmic divergences caused by
soft breaking terms at $M_{\rm GUT}$ in Supergravity
(SUGRA) models. Thus in practice  a large logarithmic  factor 
$\ln (M_{\rm GUT}^2/m_W^2)\sim 65$ is included in the decay amplitude.
\footnote{For SUSY breaking mechanisms other than gravity mediation,
such as gauge mediation, the SUSY breaking scale can be much lower than
the GUT scale, which lead to a somewhat smaller value of $\Gamma$  in
Eqn.\ref{loopdkwidth}.}

The 4-body decay mode $\tilde{t}_1 \ra b \chi_1^0 f \bar{f}'$, proceeds
through various  diagrams \cite{boehm}. An approximate  
 analytical expression for the 4-body decay amplitude
can be found in ~\cite{boehm}. However,
 the decay rate has been been calculated in \cite{boehm} without
any approximation by taking 
into account all diagrams and the interferences among them. We thank the authors
of \cite{boehm} for providing their {\tt FORTRAN} code.  
The packages {\tt SDECAY} \cite{sdecay} and {\tt CalcHEP v2.1} \cite{calchep}  are now
available for  calculating  the  decay width in the context 
of MSSM. We have cross checked the {\tt FORTRAN} code with {\tt CalcHEP v2.1}. 

In general, in any diagram  the  virtuality of the exchanged  
particle  in the propagator is important
for getting an  appreciable contribution. This gives an idea of the 
parameter space where the 4-body width is significant. 
It has been checked that for the  
$\lstop$ mass range and the model parameters used in this 
analysis  the chargino and/or the slepton 
mediated diagrams are the dominant ones  and the 4-body width is 
indeed comparable to the loop decay width.

The 4-body  and the loop decay of the $\lstop$-NLSP are of the same order in
perturbation theory, i.e. ${\cal O} (\alpha^3)$. So,  
they may compete with
each other. Our prime objective is to figure out the SUSY parameter space
where the 4-body BR may compete with or overwhelm the loop decay BR. 
In this perspective a few more points are noteworthy.

\bi
\item As discussed above  the large 
logarithmic factor in the width (see,  
Eqns.\ref{loopdkL} and \ref{loopdkLR}) presupposes the existence of a grand
unification scale. Moreover, it is also assumed that MSSM
is the correct theory all the way up to $\mgut$. 
If this is not the case  this factor can be rather small. This would enhance
the 4-body BR compared to the typical values presented in this paper.

\item If the $\lstop$ is purely right-handed,
the amplitude involves only $\Delta_R$ in Eqn.\ref{loopepsilon}.
This will suppress the amplitude for small values of the trilinear 
coupling $A_b$ (in Eqn.\ref{loopdkLR}). Additional suppression may 
come if the (common) SUSY-breaking scalar mass of the 
first two generations is large (Eqn.\ref{loopepsilon}).

\item  Any scenario with
$\tan \theta_{\til t} \simeq \Delta_L/\Delta_R$ immediately leads to a
negligible $\eps$  (Eqn.\ref{loopepsilon}) .

\item From Eqns.\ref{loopdkL} and \ref{loopdkLR} we note that  the 
mixing between $\lstop$ and $\widetilde c_L$ is  proportional
to $\tan^2\beta$. It is, therefore, relatively suppressed at 
small $\tan\beta$ which is favourable for a large 4-body BR \cite{boehm,me1}.
\ei

It has already  been pointed out that 
the $\lstop$ searches in the 4-body leptonic decay channel is especially 
promising for $3 \lsim \tb \lsim  5$ at the upgraded Tevatron \cite{me1}. 
However, it is also well known that the radiatively corrected $\mlhiggs$  
increases with $\tb$. So, what happens to the parameter spaces favourable 
for $\lstop$ searches in the 4-body decay channel if the Higgs boson 
mass bound is invoked? We are addressing this issue in this paper.

The input value of top-quark mass in this paper is important for 
two reasons. Firstly, it changes the parameter space where the  $\lstop$-NLSP  
condition is satisfied. Consequently the 4-body decay BR 
will be affected as well for a particular choice of  
SUSY model parameters. Secondly, as discussed above the,  
radiative corrections to the Higgs boson mass is very sensitive to $m_t$. 

 Before delving into the 
numerical discussions let us briefly review  the  
current status of the top mass measurement.  From 
the  published Tevatron Run-I and preliminary Tevatron Run-II data, the present 
world average for the top-quark mass is $m_t=172.7 \pm 2.9$ GeV  
\cite{massoftop}. Since there is considerable uncertainty 
in $m_t$ we shall consider in this paper two representative  
values:  
$(i)$ $m_t$=172.7 GeV 
which is the central value of the Run-I and Run-II average and 
$(ii)$ $m_t$=178.0 GeV which is on the higher side 
of the currently allowed range ( approximately 
2$\sigma$ away from the present central value ).
The second choice predicts  
relatively large $\mlhiggs$ for fixed values of other SUSY parameters and 
hence, leads to weaker constraints on the parameter space allowed 
by the bound on $\mlhiggs$.

The plan of the paper is as follows. In section \ref{numerical}, we begin 
with a discussion of the numerical tools used for calculating the sparticle 
masses and the lighter CP-even Higgs boson mass. The sub-sections \ref{msugrasec} 
and \ref{mssmsec} are devoted to the phenomenological discussions of the 
lighter top-squark 4-body decay modes respectively in mSUGRA and MSSM models  
consistent with the lighter CP-even Higgs boson mass limit from 
LEP experiments. Consequences of the theoretical 
uncertainties in the $\mlhiggs$ prediction are also analyzed. We 
shall conclude in section \ref{conclusion}.

\section{Numerical Analysis :}
\label{numerical}

We begin with representative parameter spaces in which the $\lstop$-NLSP 
condition is realized. We then draw the  
contours of its total 4-body decay BR in this parameter space. Finally 
we impose the  
lighter CP-even Higgs boson mass constraint on this parameter space. The 
sparticle spectrum in mSUGRA has been calculated 
using {\tt ISAJET v7.48} \cite{isajet}. The sparticle spectrum in MSSM has 
been calculated using our own codes. The Higgs boson mass has been 
calculated using {\tt SUSPECT v2.33} \cite{suspectv233} in both the models. 
This code uses the  full one-loop 
and dominant two-loop contributions to the self energy 
corrections in the SUSY Higgs boson mass matrix in the Dimensional 
Reduction ($\overline{DR}$) scheme \footnote{ Other public codes, 
for example, FeynHiggs \cite{feynhiggs} and CPsuperH \cite{cpsuperh} are     
also available for  calculating the properties of the Higgs sector in various 
SUSY models. 
}. 

The supersymmetric radiative corrections to the particle (top, bottom, tau) and 
sparticles masses (squarks and gauginos) have also been incorporated.  
We have set the following input values in {\tt SUSPECT v2.33} at 
the weak scale ($\mweak$): 
${1\over\alpha_{em}}$= ${1 \over 127.934}$, 
$\alpha_{s}$=0.1172, $\mtop$(pole)=172.7 or 178.0 GeV, $m_{\tau}$(pole)=1.777 GeV, 
$m_b(m_b)$=4.25 GeV, $sin^2_{\theta_W}$=0.2221. 

The $M_A$ and $\mu$ are treated as inputs in MSSM in this paper. 
All masses and mass parameters in this paper are in GeV.

The magnitude of $\mu$ can be constrained from the 
radiative electro-weak symmetry breaking (REWSB) conditions 
which produce the observed Z-boson mass. Minimizing the 
Higgs potential with radiative corrections one obtains

\begin{eqnarray}
{1\over 2}M_Z^2&=&{{m_{H_d}^2-m_{H_u}^2\tan ^2\beta }
\over {\tan ^2\beta -1}}-\mu ^2 + \Delta_R \;, \label{treemin1}
\end{eqnarray}
and
\begin{eqnarray}
\mu B &=&{1\over 2}(m_{H_d}^2+m_{H_u}^2+2\mu ^2)\sin 2\beta \;.
\label{treemin2}
\end{eqnarray},
where $m_{H_u}, m_{H_d}$ are the soft Higgs mass parameters at 
$\mweak$ and $\Delta_R$ is the radiative correction which is
given in \cite{isajet} and \cite{suspectv233} . This REWSB  
criterion restrict the $|\mu B|$ and $|\mu|$ at $\mweak$ and
leaves the $sign(\mu)$ as a free parameter in mSUGRA. One can trade the 
parameter $B$ for $\tb$ (ratio of Higgs VEVs). Eqns.\ref{treemin1} and 
\ref{treemin2} are also valid in MSSM but for arbitrary soft  masses.  

The soft SUSY breaking terms cannot be arbitrary  at $\mweak$ 
in mSUGRA.  The soft terms are renormalization group 
equation (RGE) driven  from high scale 
inputs like the common scalar mass, see sub-section 2.1. The trilinear couplings 
at $\mweak$ can also be computed  from 
a common input value at $M_{\rm GUT}$ using the RGE. The soft terms  are  
treated as independent inputs in MSSM which we shall discuss in the sub-section 2.2 .  

However, in both the models one can show that a deep charge-color-breaking (CCB) 
minima \cite{rewsb1,ccb1,ccb2} appears in the scalar potential at $\mweak$ unless

\begin{eqnarray}
|A_t|^2 \lsim 3( m_{\til {t_L}}^2+m_{\til{t_R}}^2 + m_{H_u}^2 +{\mu}^2)\nonumber\\
|A_b|^2 \lsim 3( m_{\til {b_L}}^2+m_{\til{b_R}}^2 + m_{H_d}^2 +{\mu}^2) \label{eq:ccbcondn}  \\
|A_{\tau}|^2 \lsim 3( m_{\til {\tau_L}}^2+m_{\til{\tau_R}}^2 +m_{H_d}^2+{\mu}^2 ) \nonumber 
\end{eqnarray}
where $A_t$, $A_b$ and  $A_{\tau}$ are the trilinear soft parameters of  
the top-squark, bottom-squark and tau-slepton sector respectively.  

In our analysis we have considered the ranges of model parameters such that
the CCB minima \cite{rewsb1,ccb1,ccb2} of the potential do not arise. Moreover 
all sparticle masses are required to satisfy the recent experimental bounds.  
We now make a few comments on the obtained $\lstop$ mass limits. 

The published mass limits on $\lstop$ not only depend on the
production cross-sections, they also
depend on $\Delta M = \mlstop - m_{LSP} $, where LSP can be
either $\lspone$ or $\snu$. In addition these limits depend on the 
BRs of various decay channels kinematically allowed. Moreover, the limits
from LEP also depend on $\mix$.

Most of the mass limits from 
Tevatron \cite{tevrecentstop,cdfloopfirst,ex0404028,fermistop4} and 
LEP \cite{leprecentstop,llh} have been obtained by assuming 
100\% BR of the $\loopdk$ or the $\twodk$ decay. However, the  
presence of other competent decay mode, for example the  
4-body decay mode emphasized in this paper, substantially 
changes  these limits. The CDF limit \cite{cdfloopfirst} with the 
above assumption has been included in the analysis  of 
section \ref{msugrasec}. 

Recently the D$\dzero$  collaboration \cite{dzerostop} and the ALEPH
collaboration \cite{alephstoptopsqk} put limits on $\mlstop$  
assuming complete dominance of the channel 
$\lstop \ra b \lspone \ell \nu_{\ell}$ and  
coexistence of the loop decay and the 4-body decay respectively. 
From  Fig.4b of \cite{alephstoptopsqk} the absolute 
lower limits of 
$\mlstop$ is found to be $\approx$ 63 at 95\% C.L. . This limit 
is obtained for $\Delta M  = 5 $, $\brloopdk$=22 \% and
$BR(\lstop \ra b \lspone \ell \nu_{\ell})$ =55 \% and $\mix = 56^{\circ}$.
The  $\mlstop$ limit becomes stronger when
$\brloopdk$ is increased from 22\%. It may be noted that the assumed BRs
in the above works are not realistic. For example it was shown in \cite{me1} that 
the BR of the leptonic ($\ell=e $ and $\mu$) 4-body decay mode is unlikely to  
exceed 20\% since competing hadronic 4-body decay modes are inevitably 
present. In a different approach \cite{me3} model independent 
limits on the product of branching ratios
$\branch(\lstop\ra be^{+}\nu_e \lspone)
\times \branch(\lstop^{*}\ra \bar{b}q\bar{q^{\prime}}\lspone)$, where
q and $q'$  correspond quarks of all flavours kinematically allowed, 
were obtained as a function of $\mlstop$ using Tevatron  Run-I data.
 
The LEP-II collaboration puts the limit  
$\mlsbot \gsim 96$ \cite{sbotabslower} assuming complete 
dominance of $\lsbot \ra b \lspone $ decay. The following constraints on 
sparticle masses have also been included in the analysis of 
this paper: $\mstauone \gsim 86$, $\msnu \gsim  43.7$, $m_{\wt \mu_R} \gsim  94.9$,   
$m_{\wt e_R} \gsim  99.4$ \cite{lepsusyslepton} and 
$\mchonepm \gsim  103 $ \cite{lepsusychargino}.

\subsection{The mSUGRA model:}
\label{msugrasec}

The input parameters of mSUGRA at $M_{\rm GUT}$ are the common scalar 
mass ($m_0$), the common gaugino mass ($m_{1/2}$), the trilinear scalar 
coupling term $(A_0)$, $\tan\beta$ and sign of $\mu$  
(the higgsino mass parameter)~\cite{sugrareview}. The magnitude of $\mu$ is  
fixed by the REWSB condition ~\cite{rewsb1,rewsb} and depends on the input top 
quark mass\footnote{The top-quark mass is very crucial for calculating 
$m_{H_u}$ and $\Delta_R$ in Eqns.\ref{treemin1} and \ref{treemin2}. }.
 
Since the NLSP nature of the 
$\lstop$ is very crucial for our work, we 
have demarcated different regions in the $\azero - \tb$ parameter space 
where $\lstop$ becomes the NLSP. A detailed discussion is given in 
sec.II of \cite{me1} (see Fig.1 and 2 \footnote{One can  
easily find $\mlspone$ and $\mchonepm$ from the parameter set used 
in the figures.} 
of \cite{me1}), where it was assumed that $m_t$=175.0 . Similar information 
is given in Fig.\ref{fig:mzmhfmupmt1727} and 
Fig.\ref{fig:mzmhfmupmt178} in this paper 
for $m_t$=172.7 and $m_t$=178.0 respectively.  
Several contours of the combined BR of all 4-body decay modes 
are also shown. In order to identify the parameter space allowed by the  
$\mlhiggs$ constraints, we present the contours of $\mlhiggs \simeq $111.4 (the  
weaker bound) in both the figures  and $\mlhiggs \simeq $114.4 (the 
experimental bound) in the right panel of Fig.\ref{fig:mzmhfmupmt178} only.

The choice of other mSUGRA input parameters 
used are given in the figure captions. These choices are guided by the 
following facts. If $\mzero >> \mhalf$, the gauginos are naturally 
light. This disfavours the $\lstop$-NLSP scenario. For 
$\mhalf >> \mzero $, on the other hand, one may require 
large $|\azero|$ for this scenario. However,  
such high values  of $|\azero|$ may   
violate of the CCB conditions, see Eqn.\ref{eq:ccbcondn}.
So, in our analysis we focused on $\mzero \sim \mhalf$ regions for mSUGRA. 
However, the $\lstop$-NLSP is quite common in larger regions in the 
parameter space of the unconstrained MSSM since the 
soft breaking parameters are arbitrary. This model will be analyzed in the 
next section.

For each $\tb$ there is a range of $|\azero|$ 
( ${|\azero|}_{min} < |\azero| < {|\azero|}_{max}$ ) at $M_{\rm GUT}$ which 
corresponds to the 
$\lstop$-NLSP (see Fig.\ref{fig:mzmhfmupmt1727} and 
\ref{fig:mzmhfmupmt178}). Otherwise $\stauone$ or $\lsptwo$ or 
$\chonepm$ happens to be the NLSP and 
$\lstop$  decays into other  
 2-body or 3-body decay channels. 

For $|\azero| > {|\azero|}_{max} $,  
$\lstop$ becomes the LSP which is forbidden. On the other hand 
if $|\azero| < {|\azero|}_{min} $ for the given set of mSUGRA parameters, 
$\lstop$ becomes heavier than the lighter chargino and the 
$\lstop$-NLSP condition is violated.  We, therefore, concentrate on 
$|A_0|\sim {|\azero|}_{min} $ so that the
chargino virtuality is small \cite{boehm,me1} and consequently  
the 4-body BR is appreciable. 
The off-diagonal terms in the top-squark mass matrix  
have the form ($A_{t}$ -$\mu\over\tb$).  So as 
long as $A_{t}$ at $\mweak$ and $\mu$ 
have opposite signs the NLSP scenario is favored.

For $m_t$=172.7, the $\lstop$-NLSP is realized in a large region of the 
parameter space (see Fig.\ref{fig:mzmhfmupmt1727}). As expected the total 
BR of the 4-body decays is appreciable for relatively low 
value of $\tb$ \cite{boehm,me1}. The region above  the $\mlhiggs 
\simeq 111.4$ contour cannot be excluded due to the theoretical
uncertainties mentioned in the introduction. It should be noted that 
even with the weaker bound on $\mlhiggs$ the total 4-body decay BR is 
 $\lsim$ 5\% for $\mzero=200$, $\mhalf=145$ (see the left panel of  
Fig.\ref{fig:mzmhfmupmt1727}). For $\mzero=140$, $\mhalf=180$, on 
the other hand, the total 4-body decay BR can be as large as  40\%   
(see the right panel of 
Fig.\ref{fig:mzmhfmupmt1727}). However, the whole 
parameter space in Fig.\ref{fig:mzmhfmupmt1727} is disfavoured if 
the experimental bound is imposed.

A similar analysis for $m_t$=178.0 is shown in 
Fig.\ref{fig:mzmhfmupmt178}. In  the left panel of
Fig.\ref{fig:mzmhfmupmt178} with
 $\mzero=200$, $\mhalf=145$ we have some allowed region if 
the weaker bound on $\mlhiggs$ is imposed. Still the total 4-body decay  
BR can be at most 20\%. In the right panel of Fig.\ref{fig:mzmhfmupmt178} with
 $\mzero=140$, $\mhalf=180$ we have some allowed region even if 
the experimental bound is imposed. However, 
the total  4-body decay  BR can be at most 
20\%. Of course with the weaker bound larger 4-body decay BRs ( 
$\lsim$ 60\%) are allowed.
It should also be noted that for a fixed $\mzero$, $\mhalf$ cannot 
be increased arbitrarily because that would lead to other instabilities 
of the scalar potential \cite{rewsb1,ccb1,ccb2,ufbas}.

We have shown the variation of $\mlstop$($\mlhiggs$) as a
function of $\azero$ in the 
left (right) panel of Fig.\ref{fig:mz140mhf180mhmst} for $\mtop$=178.0 
for three values of $\tb$. Other model parameters are given in the 
figure caption. The 
$\lstop$-NLSP criterion is not imposed 
in this particular figure. However, the range of 
$\mlstop$ satisfying the $\lstop$-NLSP condition can be easily 
found out by comparing with the right 
panel of Fig.\ref{fig:mzmhfmupmt178}. Then one can read off  
the lighter chargino and neutralino masses from the left panel of 
Fig.\ref{fig:mz140mhf180mhmst}.  A $\lstop$ having mass in the range 
shown in Fig.\ref{fig:mz140mhf180mhmst} may be copiously produced at 
Tevatron Run-II. From the right panel it is clear that  
$\tb$=5 is marginally allowed only if the weaker bound on  
$\mlhiggs$ is required. On the other hand if  $\tb\gsim$ 9  
then $\mlhiggs$ is consistent  even with the experimental 
bound, but the  total 4-body decay BR is negligible as can be seen 
from the right panel of Fig.\ref{fig:mzmhfmupmt178}. 

The scatter plots in Fig.\ref{fig:tb59a630mup178}  
demarcate the $\lstop$-NLSP region consistent 
with the bound on $\mlhiggs$ in the 
$\mzero-\mhalf$ parameter space for positive $\mu$, $m_t$=178.0, 
$\tb$=5 (left panel) and $\tb$=9 (right panel).  
In the left panel of Fig.\ref{fig:tb59a630mup178} with 
$\azero$=-630 we have some region allowed by the weaker bound on 
$\mlhiggs$. The total 4-body decay  BR can be as large as 50\%. In the 
right panel of Fig.\ref{fig:tb59a630mup178} for $\tb$=9 we have some 
allowed region even after imposing the experimental bound on $\mlhiggs$ 
but the total 4-body decay BR can be at most 10\%. With the weaker bound on   
$\mlhiggs$ the whole $\lstop$-NLSP region is allowed. In 
the regions with $\mlhiggs \simeq$ 111.4 the total 
4-body BR is $<$5\%.  

We have repeated the analysis for  $m_t$=172.7 using the same
parameters as in Fig.\ref{fig:tb59a630mup178}. In this 
case  the entire  $\lstop$-NLSP region in the $\mzero-\mhalf$ parameter 
space  is  disfavoured for $\tb$=5 even for the weaker bound on  
$\mlhiggs \gsim$ 111.4. For $\tb$=9 on the other hand almost the entire
parameter space is allowed by the weaker bound on $\mlhiggs$.  
However, the total 4-body BRs can 
be at most 10\% for $\tb$=9.

In \cite{me1}, we have studied  the $\lstop$ search prospects at 
Tevatron Run-II using the one lepton with 2 or more jets accompanied by a 
large amount of missing energy signal assuming 
$m_t$=175. This signal topology  
arises from the pair production of the lighter top  
squark followed by the leptonic 4-body  decay of one lighter top-squark  
($\fourdklep$) while the other decays hadronically ($\fourdkhad$).  
We have shown that, $\brfourdklep \approx$ 20\%, where 
$\ell$ = e and $\mu$ (see Figs.3 and 4 of \cite{me1} ), 
is adequate for detection at Tevatron Run-II. From the present analysis we 
find that such large leptonic BRs cannot be realized at least in the  
mSUGRA model for  
$m_t$=172.7 and not even with  the rather conservative choice $m_t$=178.0 
irrespective of the $\mlhiggs$ bound invoked. Our analysis 
also show that the $\lstop$ mass 
limits from LEP and Tevatron Run-I assuming 100\% BR of the loop decay 
of $\lstop$ are approximately valid in 
the context of mSUGRA if $m_t$ is close to its current 
central value (172.7) irrespective of the uncertainties in the 
predicted $\mlhiggs$.  However, in view of the uncertainties in $m_t$ and 
in the calculated $\mlhiggs$ one cannot conclusively
exclude the possibility of 
a relatively large total BR of all 4-body decay modes (see, $\eg$, the right 
panel of Fig.\ref{fig:mzmhfmupmt178}).  

We have shown some points in the parameter space excluded  
by the CDF limits on $\mlstop$ in   
Fig.\ref{fig:mzmhfmupmt1727} (left panel), Fig.\ref{fig:mzmhfmupmt178} 
(left panel) and Fig.\ref{fig:tb59a630mup178} using the 
heavy dotted squares. For each of 
these points  $\mlspone \simeq $ 51, $\mlstop \lsim 102 $ and 
4-body BR $\lsim$ 10\%. It is heartening to note that even the modest 
statistics from Tevatron Run-I can already put constraints on $\azero$ 
which is rather difficult to constrain from other measurements. However, 
the CDF excluded points are already ruled even by the weaker bound 
$\mlhiggs \gsim$ 111.4 in the context of mSUGRA. One notable 
exception is found in the right panel of Fig.\ref{fig:tb59a630mup178}. Here 
we find that a small region of parameter space allowed  by the weaker 
bound on $\mlhiggs$ is ruled out by the CDF constraint.

\subsection{The MSSM model:}
\label{mssmsec}

In the MSSM with a universal gaugino mass, the SU(2) gaugino mass 
parameter ($M_2$), $\mu$ and $\tb$ completely describe the 
neutralino and chargino masses and mixings. Depending on the 
relative magnitudes of $M_2$ and $\mu$, the model is  
categorized  in three different scenarios. They are the higgsino 
($M_2 >> \mu$), the gaugino ($M_2 << \mu$) and the mixed 
($M_2 \approx \mu$)scenarios. The lighter chargino ($\tilde\chi_1^{\pm}$) 
and the two lighter neutralinos ($\tilde\chi_1^0$ and $ \tilde\chi_2^0$)  
all have approximately the same mass ( $\approx \mu$) in the higgsino scenario. Thus 
it is difficult to accommodate the $\lstop$-NLSP without fine adjustments of 
the parameters. The $\lstop$ happens to be the NLSP in  large regions  of the 
parameter space in the gaugino and the mixed scenarios.  In this  
paper we studied  the mixed scenario. However, the main 
results and conclusions are qualitatively valid in the gaugino scenario. 

To calculate $\mlstop$ and $\mlhiggs$ in MSSM, we use the
parameter $\xr$ (instead of $\azero$) as input, where  
\br
\xr= {(A_{t} - \mu\cb)\over {\sqrt {\msquleft\msqurht}}}, 
\label{xrmssm}
\er
$\msquleft$ and $\msqurht$ are the soft SUSY breaking 
mass parameters for the first generation left and 
right-squark respectively at the weak scale.

In our analysis the first generation left and right-squark 
soft masses are assumed to be the same, $\ie$, 
$m_{\wt{u}_L}=m_{\wt{u}_R}= 
m_{\wt{d}_R} \equiv m_{\wt{q}}$. The charged
slepton sector is also treated similarly, $\ie$,
$m_{\wt{e}_L}=m_{\wt{e}_R} \equiv m_{\wt{l}}$, but the universality of
squark and slepton masses is not assumed. These mass patterns can 
be realized in some variations of the mSUGRA model.

For the third generation squarks the above   
equality may not hold even in mSUGRA due to, for example, RGE driven 
by the Yukawa terms. However, the $SU(2)_L$ invariance  require 
$m_{\wt{b}_L}$=$m_{\wt{t}_L}$ and  $m_{\wt{\nu}_L}$=$m_{\wt{\ell }_L}$, neglecting 
the D-term contributions.  In this paper the right and left-squarks soft 
masses of the third generation are  assumed to be equal to $\msq$. The  physical 
masses are calculated after incorporating the corresponding fermion masses, the 
D-terms etc in the mass matrix. 

For a given $\mu$, $\tb$, $\msq$ and $\xr$, the trilinear soft 
breaking mass parameters ($A_t$) of the top-squark sector 
can be calculated from Eqn.\ref{xrmssm}. The same  
parameter for bottom-squark ($A_b$) and tau-slepton ($A_{\tau}$) sectors 
are treated as free input parameters. Using these inputs the 
third generation squark and slepton masses and mixing angles can be calculated. 

We have used the CP-odd Higgs boson mass ($M_A$) and $\mu$ as the free 
parameters in {\tt SUSPECT v2.33} to calculate the lighter 
CP-even Higgs boson mass. 

The off-diagonal terms in the top-squark mass matrix have the 
form $m_t$$\xr$$\msq$. 
So, the masses and the mixing angle of the top-squark sector are 
controlled by these parameters. For negative $\xr$, the 
$\mlstop$ and $\mhstop$ 
are exactly the same as that for positive $\xr$. The mixing angle  
 depend on the relative magnitude of the  diagonal and off-diagonal 
terms in the top-squark mass matrix.

The $\mlstop$ ($\mlhiggs$) as a function of $\xr$ has been shown 
for $m_t$=172.7 in the left (right) panel of Fig.\ref{fig:xrtbmstmh} for 
four values of $\tb$ (see the figure caption   
for other input MSSM parameters). $\mlstop$ is insensitive  
to $m_t$ and $\tb$. 

The lighter CP-even Higgs boson mass has a periodic dependence 
on $\xr$. For a given $\xr$, if $\tb$ increases the Higgs boson mass also 
increases. If $\tb$=5 (15) and $\mtop$=172.7, $\mlhiggs \simeq$ 114.4 
for $\xr \simeq$2.0 (1.5) and $\mlhiggs \simeq$ 111.4 for 
$\xr \simeq$ 1.75 (1.30). It is clear from  Fig.\ref{fig:xrtbmstmh} that
both the $\lstop$-NLSP condition and the experimental bound on $\mlhiggs$  
may be satisfied  for large positive values of $\xr$. In our subsequent 
analyses we shall only consider such values of  $\xr$.

Due to the large number of free parameters in  MSSM we 
varied two important parameters at a time  
keeping the rest fixed.

\vskip0.3cm
$\bullet$ \underline{$\xr - \tb$}
\vskip0.3cm

We have demarcated the $\lstop$-NLSP regions in 
$\xr - \tb$ plane in the 
left (right) panel of Fig.\ref{fig:xrtanbmtwo300msl250} 
for $m_t$=172.7 (178.0). Several contours for fixed values of the total
BR of 4-body decays are also shown. 
Comparing with Fig.\ref{fig:xrtbmstmh} 
we find that the mass of the $\lstop$-NLSP is 
kinematically accessible at Tevatron Run-II.
For each $\tb$ there is a range of $\xr$ 
( ${|\xr|}_{min} < |\xr| < {|\xr|}_{max}$ ) satisfying the $\lstop$-NLSP
condition ( recall the $\azero$ ranges in Figs \ref{fig:mzmhfmupmt1727} 
and \ref{fig:mzmhfmupmt178} ).  

From each BR contour it follows that 
 when $\xr$ is increasing smaller values of $\tb$  are required. For 
large $\xr$, $\mlstop$ decreases and the virtuality of the exchanged
chargino and/or slepton becomes large and tends to suppress the 4-body
decay width. In order to keep the BR fixed $\tb$ must be correspondingly 
reduced.

We have also plotted several contours for  
$\mlhiggs \simeq 114.4$ and a few other Higgs boson masses. Comparing the 
left panel of Fig.\ref{fig:xrtanbmtwo300msl250} with  
the right panel of Fig.\ref{fig:xrtbmstmh} we find that 
for  $\mlhiggs \simeq $ 111.4 the $\lstop$-NLSP condition is violated.

As has already been noted, for $m_t$=178.0 the constraints imposed on the
parameter space by the bound on $\mlhiggs$ are less severe 
(see the right panel of Fig.\ref{fig:xrtanbmtwo300msl250}).
The total 4-body BR can be as large as 90\% for $\tb \simeq 9$ for
both choices of $m_t$ in Fig.\ref{fig:xrtanbmtwo300msl250}. Even a
t $\tb \simeq 24$ the total 4-body BR is small but not 
negligible ($\approx$ 10\%). In this case it will be interesting to 
search for hadronic 4-body decays of $\lstop$ at the Large Hadron Collider (LHC) or  
the International Linear Collider (ILC). Even if $\tb \simeq $3,  
$\mlhiggs \simeq$114.4 is allowed. Consequently almost the  
entire  parameter space in the right panel of Fig.\ref{fig:xrtanbmtwo300msl250}  
is allowed. Such low values of $\tb$ naturally guarantees large total 
BR of the 4-body decay modes. Moreover  we find that the BR of 
leptonic 4-body decays can be as large as 20\%, required for an 
observable signal at Tevatron Run-II \cite{me1}.

It is expected that the upcoming collider searches will either 
discover the Higgs boson or push its mass bound upward. Moreover, the   
theoretical uncertainties on the Higgs boson mass is likely to be 
significantly reduced ($\delta \mlhiggs \lsim $ 0.5) \cite{sven1,sven3}.  We 
would now like to comment on the impact of such possible improvements in
our understanding of $\mlhiggs$ on $\lstop$ searches. For illustration  
we have plotted  contours of $\mlhiggs \simeq $ 116.0, 118.0 and 119.0 
(118.0, 120.0 and 122.0) for $m_t$=172.7 (178.0) in the left (right) panel of 
Fig.\ref{fig:xrtanbmtwo300msl250}. The total 4-body 
BR $\gsim$ 30\% can be completely (nearly) ruled out by an improved 
Higgs boson mass bound, $\mlhiggs \gsim $ 119.0 (122.0) if $m_t$=172.7 (178.0) 
for our choice of other SUSY parameters.

\vskip0.3cm
$\bullet$ \underline{$\xr - M_A$}
\vskip0.3cm

We have shown the $\lstop$-NLSP regions, a few 4-body BR and 
$\mlhiggs$ contours in the $\xr - M_A$ plane for $m_t$=172.7 and 
$\tb$=5 (9) in the left (right) panel of Fig.\ref{fig:xrma178tb59mup}. The 
other MSSM model parameters used are  mentioned in the figure caption.  

The lighter top-squark mass is almost independent of $M_A$ if the 
other MSSM parameters are fixed. So, the $\lstop$-NLSP region remains the same 
when $M_A$ is varied for fixed $\xr$. The input value of $M_A$ 
can affect the CCB condition (see Eqns.\ref{ewsbmssm}, \ref{treemin1} 
and \ref{eq:ccbcondn}) through the soft breaking Higgs boson masses. This 
disallows some regions of the parameter space for large $\xr$. However, 
the CP-even Higgs boson mass depends  on $M_A$. On the  
other hand $M_A$  has a nontrivial impact on the loop decay 
width, see, Eqns.\ref{loopdkL} and \ref{ewsbmssm} or Ref.\cite{hikasa}.  
For a fixed $\xr$ if $M_A$ increases the loop decay width also increases. 
However, the  4-body decay width is insensitive to $M_A$. So, for 
large $M_A$ the total 4-body BR  decreases for a given 
$\xr$. This is very transparent in Fig.\ref{fig:xrma178tb59mup} for 
relatively large values of $\xr$ corresponding to large virtuality of the 
exchanged particles in the 4-body decay amplitude. 

We find that the total 
4-body BR (leptonic BR) can be as large as 90\% (20\%) in 
significant regions of the parameter space
 for $\tb$=5. If the experimental bound on the 
Higgs boson mass is invoked then the total 4-body 
BR (leptonic BR) $\simeq$ 90\% (20\%) is  
allowed if $M_{A}\gsim$ 300 (400) and $\xr \lsim$  2.17 (2.08)
for $\tb$=5 (see the left panel of Fig.\ref{fig:xrma178tb59mup}). 
We have shown $\mlhiggs \simeq$ 111.4 contour in the 
left panel of Fig.\ref{fig:xrma178tb59mup}. It is clear that 
almost entire parameter space is allowed by  
the weaker bound on the Higgs boson mass. It has also been checked  
that the whole $\lstop$-NLSP region is disfavoured for  
$\mlhiggs \gsim $117.4. 

For $\tb$=9, in the entire $\lstop$-NLSP region in 
Fig.\ref{fig:xrma178tb59mup} corresponds to 
$\mlhiggs$ $\gsim$ 114.4. However, the total 
4-body BR (leptonic BR) $\simeq$ 90\% (20\%) occurs only in restricted 
regions, $\ie$, $M_{A} \lsim 500 $ and $\xr \lsim$ 2.025 as can 
be seen from the right panel of Fig.\ref{fig:xrma178tb59mup}. 

Since the lower bound on  Higgs boson mass 
may improve in future, we have shown  
several $\mlhiggs$ contours, $\eg$, 115.0 and 115.7 (116.0, 117.0, 117.4, 
117.8, 118.4 and 119.0 ) for $\tb$=5 (9) in the left (right) panel 
of Fig.\ref{fig:xrma178tb59mup} to illustrate the consequences 
for the BR under study. 

\section{Conclusions}
\label{conclusion}
\vspace*{5mm}

The lighter top-squark turns out to be the NLSP in a  significant region 
of the supersymmetric parameter space both in 
constrained and unconstrained  models like mSUGRA and MSSM respectively. 
The $\lstop$-NLSP has 
two competing decay modes for the top-squark masses within the reach  
of  the Tevatron Run-II. They are the loop induced $\loopdk$ and 
the 4-body decay, $\fourdk$. These two decay modes may have comparable 
decay widths over  large regions of parameter spaces in both the 
models \cite{boehm,me1}. However, for low $\tb$ the 
loop decay width is suppressed which in turn enhances 
the 4-body decay BRs. On the other hand the current bound 
$\mlhiggs \gsim $ 114.4 from LEP disfavours low values of 
$\tb$. It is therefore important to identify the parameter spaces
allowing large total BRs of the 4-body decays after imposing 
the $\mlhiggs$ constraint. 

 The theoretical uncertainties in the calculated $\mlhiggs$ using any 
currently available code is $\approx$ 3
due to the yet unknown higher-order corrections. In view of this we have 
assumed  that realistically the bound on the calculated  $\mlhiggs$ can 
be relaxed to $\mlhiggs \gsim 111.4$.

The  masses of $\lstop$ and $\mlhiggs$ crucially depend on the  
input value of $m_t$.  We have considered the current world 
average central value of $m_t$ which is $m_t$=172.7 as well as 
 $m_t$=178.0, which is approximately 2$\sigma$ away 
from the central value and yields somewhat weaker  constraints. 

\begin{figure}[htb]
  \begin{center}
   \begin{tabular}{cc}
\hspace*{-3.3cm}   \mbox{\epsfxsize=0.8\textwidth
      \epsffile{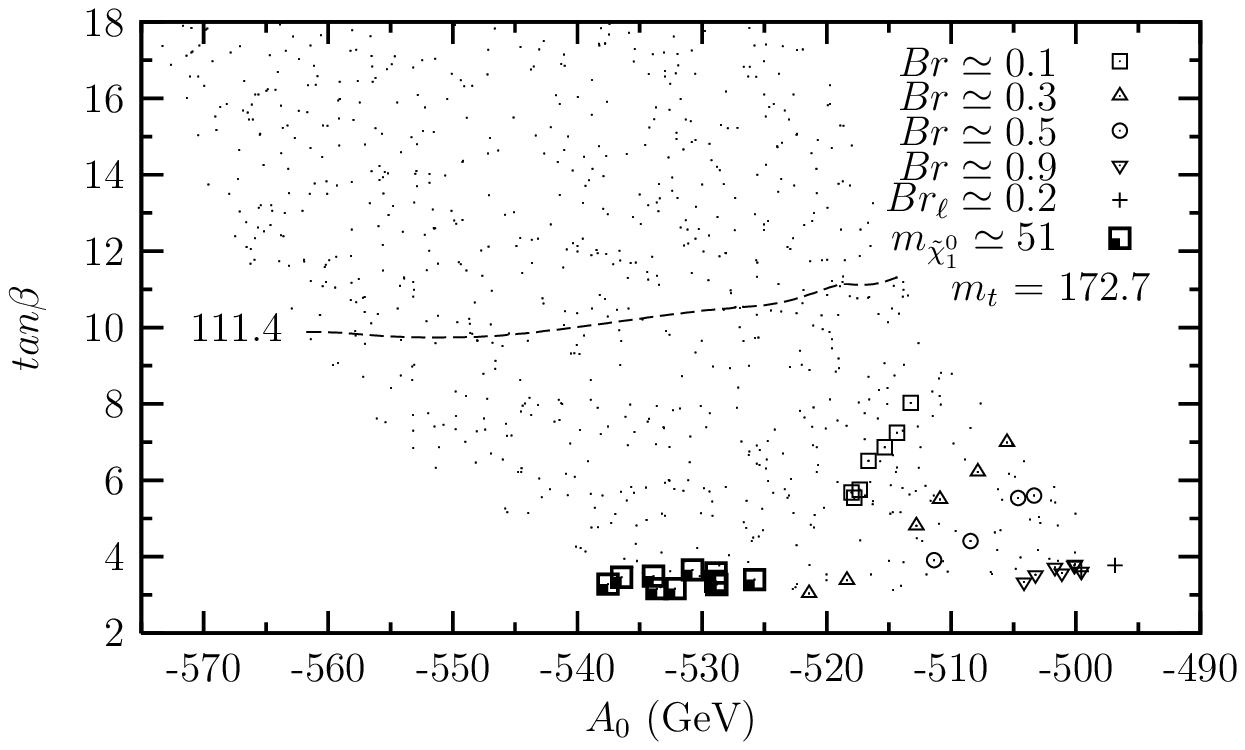}}
   &
\hspace*{-5.1cm}  \mbox{\epsfxsize=0.8\textwidth
      \epsffile{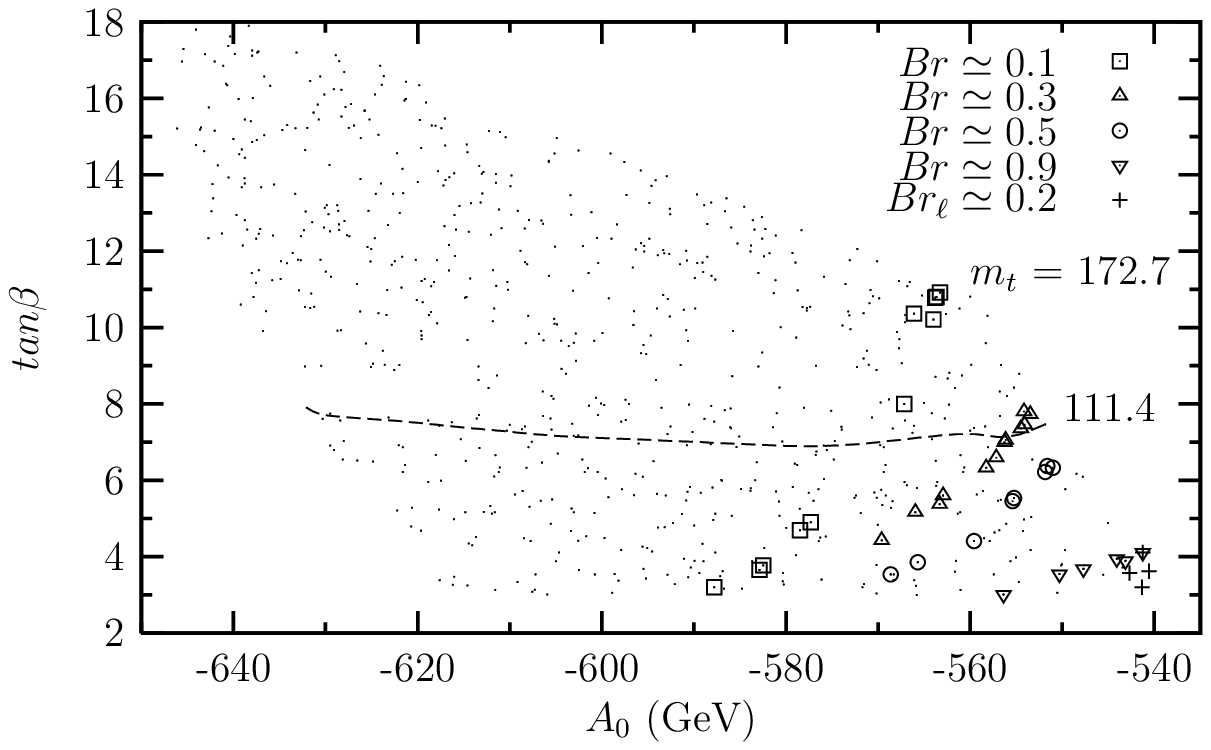}}
  \end{tabular}
  \end{center}
\vspace{-11.4cm}
\caption{ \label{fig:mzmhfmupmt1727}
{\small
The $\lstop$ becomes the NLSP in the whole marked  region 
in the $\azero -\tb $  plane in the mSUGRA model for $\mzero=200(140)$, 
$\mhalf=145(180)$, $\mu > 0$ and $m_t$ =172.7 in the 
left (right) panel. The bounds on sparticle masses and other 
SUSY constraints, see the text, are also imposed.  
The CP-even Higgs boson mass contour for 111.4 is shown and the region 
above the contour is allowed (see the text for a brief review of the 
theoretical uncertainties in the calculated $\mlhiggs$).           
The whole marked region is ruled out by the experimental 
bound  $\mlhiggs \gsim$ 114.4. Several contours of total 4-body decay BR  
lighter top-squark are shown by different point styles, see the legends. 
The `{\bf \small +}' shows the regions where $\brfourdklep \simeq 20\%$ 
for $\ell$=e and $\mu$. The regions marked with heavy dotted square 
box are excluded by  the  CDF limit ( $\mlstop \gsim$ 102 GeV for 
$\mlspone \approx $ 51 GeV) obtained by assuming 100\% BR of the loop 
decay.
}}
\end{figure}

\begin{figure}[htb]
  \begin{center}
   \begin{tabular}{cc}
\hspace*{-3.3cm}   \mbox{\epsfxsize=0.8\textwidth
      \epsffile{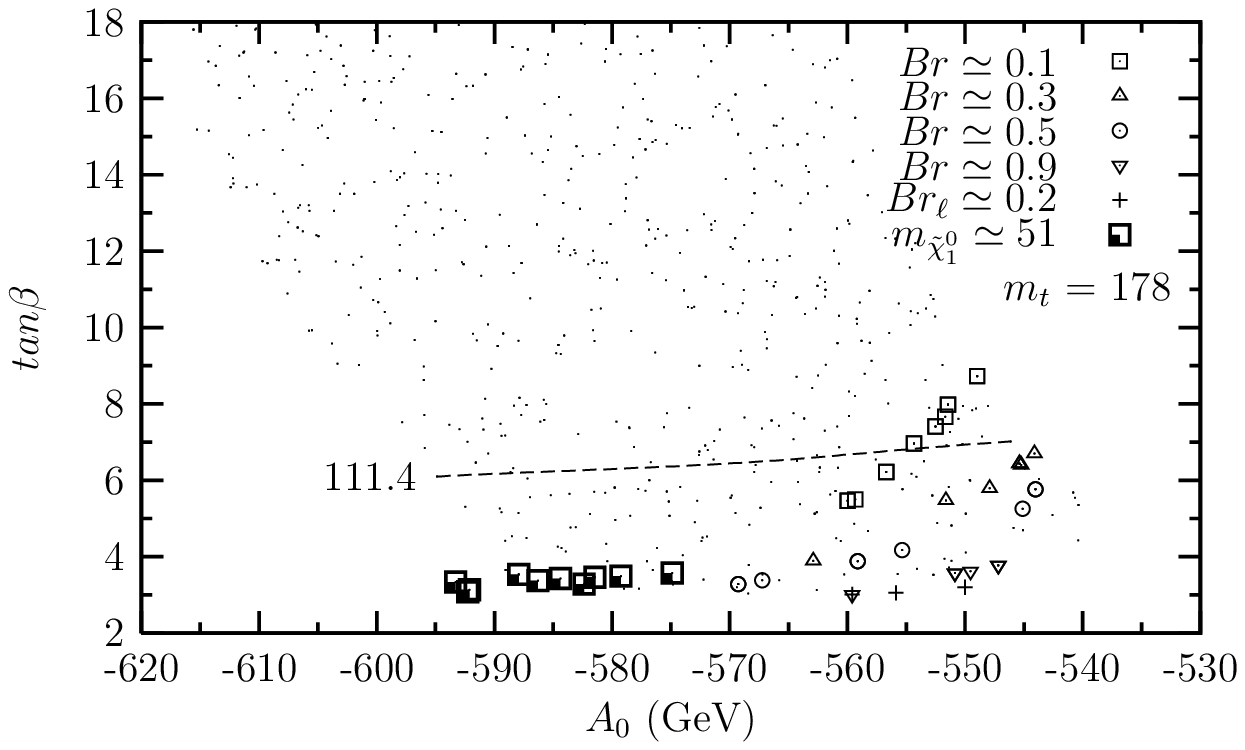}}
   &
\hspace*{-5.1cm}  \mbox{\epsfxsize=0.8\textwidth
      \epsffile{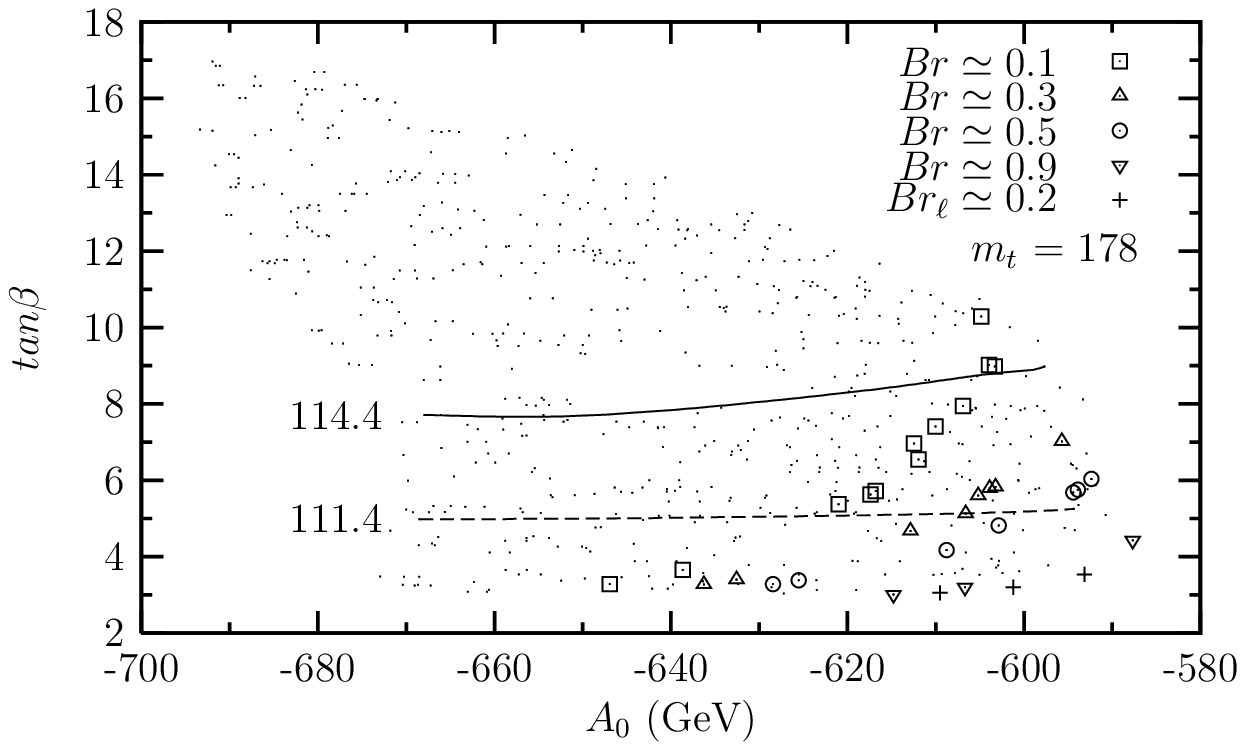}}
  \end{tabular}
  \end{center}
\vspace{-11.4cm}
\caption{ \label{fig:mzmhfmupmt178}
{\small
The $\lstop$ becomes the NLSP in the whole marked  region 
in the $\azero -\tb $  plane in the mSUGRA model for $\mzero=200(140)$, 
$\mhalf=145(180)$, $\mu > 0$ and $m_t$ =178.0 in the 
left (right) panel. The bounds on sparticle masses and other 
SUSY constraints, see the text, are also imposed.  
 The CP-even Higgs boson mass contours for 111.4 (111.4 and 114.4) 
are plotted in the left (right) panel and the regions above the 
contours are allowed from the respective bounds.
The whole marked region in the left (right) 
are ruled out from $\mlhiggs \gsim$ 114.4 (117.4) (see the text).  
The total 4-body decay BR's contours of 
lighter top-squark are shown by different point styles, see the legends. 
The `{\bf \small +}' shows
the regions where $\brfourdklep \simeq 20\%$ for $\ell$=e and $\mu$.  
The regions marked with heavy dotted square box can be excluded 
by the  CDF limit (see Fig.\ref{fig:mzmhfmupmt1727} caption) . 
}}
\end{figure}

\begin{figure}[htb]
  \begin{center}
   \begin{tabular}{cc}
\hspace*{-3.3cm}   \mbox{\epsfxsize=0.8\textwidth
      \epsffile{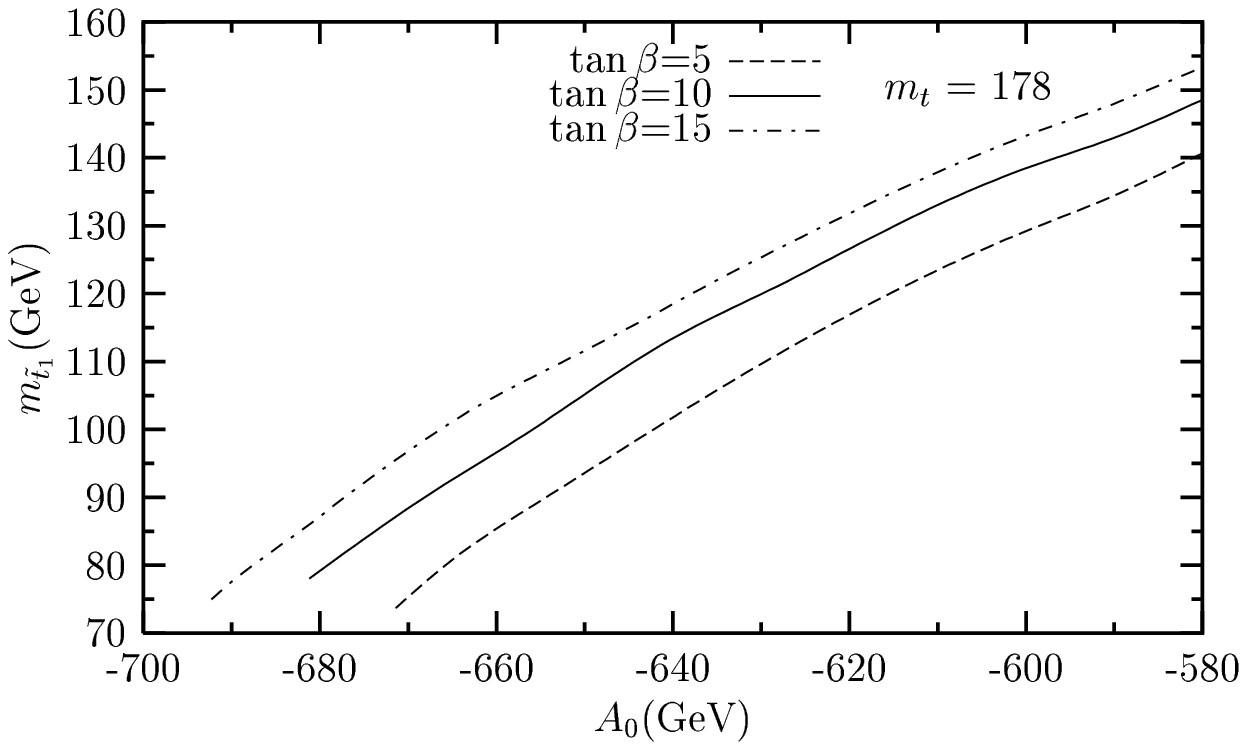}}
   &
\hspace*{-5.1cm}  \mbox{\epsfxsize=0.8\textwidth
      \epsffile{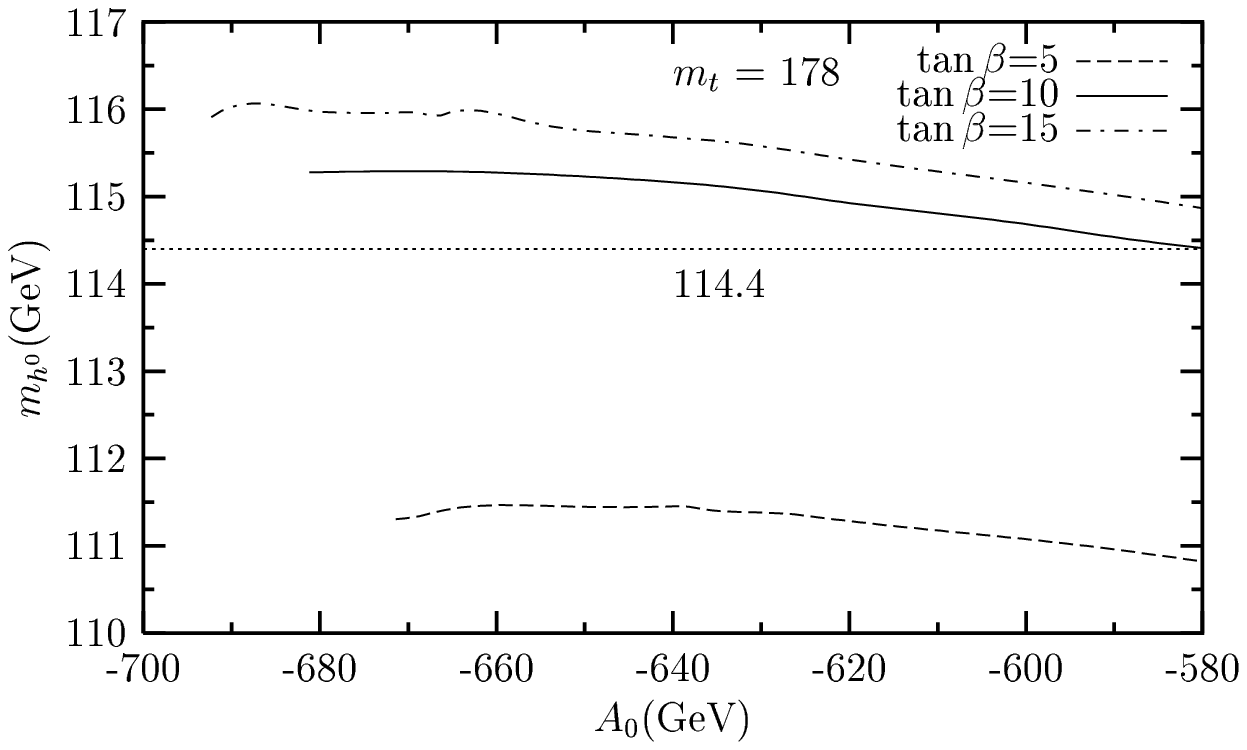}}
  \end{tabular}
  \end{center}
\vspace{-11.4cm}
\caption{ \label{fig:mz140mhf180mhmst}
{\small
The $\mlstop$ ($\mlhiggs$) as a function of $\azero$ for 
for $\mzero=140 $, $\mhalf=180$, $\mu > 0$ and $\tb$=5,10,15 in the 
left (right) panel in mSUGRA for $m_t$=178.0. The $\lstop$-NLSP criterion  
is relaxed in this particular figure.   
}}
\end{figure}

\begin{figure}[htb]
  \begin{center}
   \begin{tabular}{cc}
\hspace*{-3.3cm}   \mbox{\epsfxsize=0.8\textwidth
      \epsffile{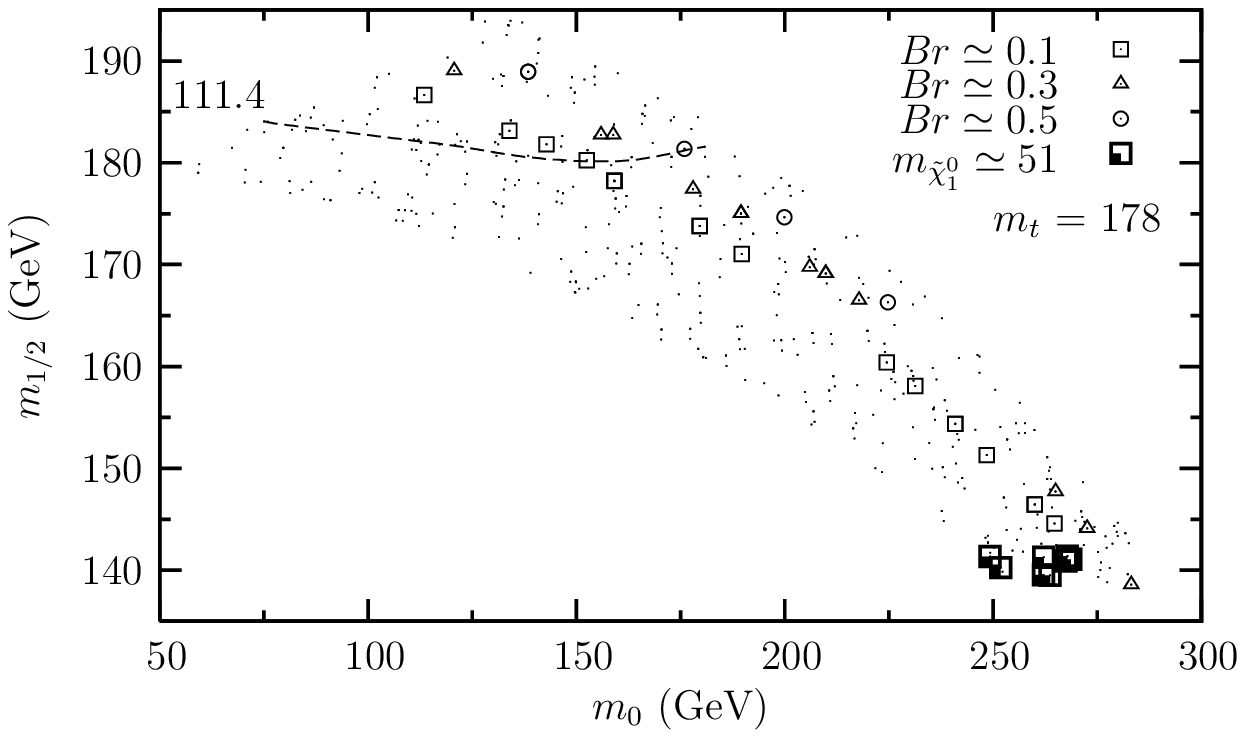}}
   &
\hspace*{-5.1cm}  \mbox{\epsfxsize=0.8\textwidth
      \epsffile{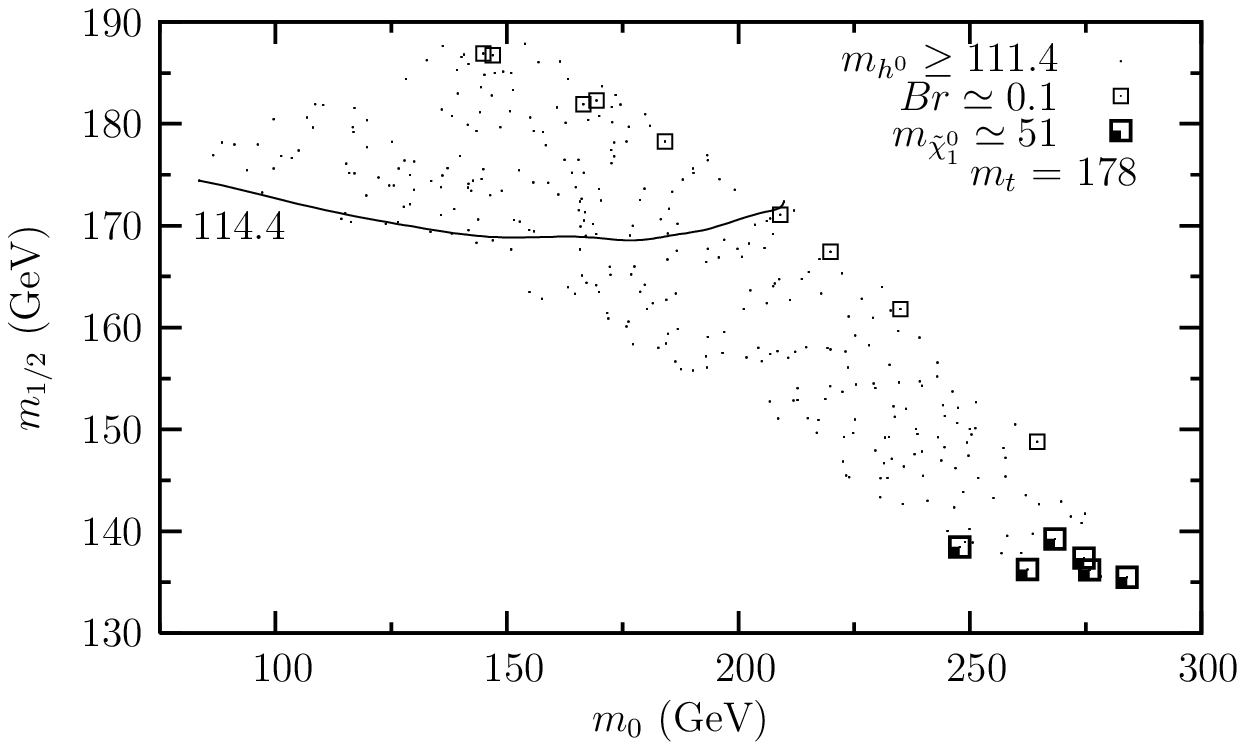}} 
  \end{tabular}
  \end{center}
\vspace{-11.4cm}
\caption{ \label{fig:tb59a630mup178}
{\small
The $\lstop$ becomes the NLSP in the whole marked  region 
in the $\mzero -\mhalf $ plane in the mSUGRA model for $\azero=-630$,
 $\mu > 0$, $m_t$=178.0 and $\tb$ =5(9) in the left (right) panel. 
The bounds on sparticle masses and other 
SUSY constraints, see the text, are also imposed.   
The CP-even Higgs boson mass contour for 111.4 (114.4) 
is plotted in the left (right) panel. The $\mlhiggs$ happens to be 
$\gsim$ 111.4 in the whole marked region in the right panel.  The regions 
above the  contour are allowed from the respective bound on $\mlhiggs$. The 
whole marked regions in the left (right) panel are ruled out from 
$\mlhiggs \gsim$ 114.4 (117.4). The  contours of total 4-body decay BR of the 
lighter top-squark are shown by different point styles, see the legends. 
The   
regions marked with heavy dotted square box can be excluded 
by the  CDF limit (see Fig.\ref{fig:mzmhfmupmt1727} caption).  
}}
\end{figure}

 We have demarcated the $\lstop$-NLSP regions in the mSUGRA model for 
$\mtop$=172.7 and 178.0 in the $\azero-\tb$ ( Figs.\ref{fig:mzmhfmupmt1727} 
and \ref{fig:mzmhfmupmt178})  and $\mzero-\mhalf$ planes 
(Fig.\ref{fig:tb59a630mup178}). We found that for a given $\tb$, a limited range of 
$\azero$ with $\azero \sim |\azero|_{min}$ allows large 4-body BRs.
For demonstration we have presented  in the $\azero - \tb$ plane
a few 4-body BR contours for two representative 
choices of $\mzero$ and $\mhalf$ (see sub-section \ref{msugrasec} and the  
figure captions). For $m_t$= 172.7 all parameter spaces with the total BR 
of the 4-body decays $>$ 10 (40)\% are disfavoured even if the weaker Higgs boson 
mass bound  is invoked (see the left (right) panel of 
Fig.\ref{fig:mzmhfmupmt1727}). The experimental bound on $\mlhiggs$ rules out 
any parameter space with numerically significant BR of the above 
decays. For $m_t$= 178.0 the experimental  
bound on $\mlhiggs$ leaves no room for total BR$>$ 10 (20)\% 
(see the left (right) panel of Fig.\ref{fig:mzmhfmupmt178}). The weaker  
bound, however, allows BRs as large as 30 - 50 \% in a small region of  
the parameter space (the right panel of the above figure).  
  
Complementary information in the  $\mzero -\mhalf$ plane, for 
$\azero$=-630, sign$(\mu)>0$ and 
$m_t$=178.0 are presented in Fig.\ref{fig:tb59a630mup178}. For $\tb$=5 the 
weaker  
constraint on $\mlhiggs$ allows BRs as large as 60\% (see the left 
panel) in a small parameter space, while the experimental bound on 
$\mlhiggs$ practically rules out the possibility of a sizable BR of
the decay channels of interest.  For $\tb=9$, the total
4-body BR can be at most 10\% due to  the experimental bound on Higgs boson mass. 

We have also excluded small regions in $\azero-\tb$ and 
$\mzero -\mhalf$ plane from the CDF limits on the mass of
$\lstop$ in the jets+$\met$ channel. In most cases, however,
these constraints are superseeded even by the weaker bound on 
$\mlhiggs$.

In conclusion we note that in mSUGRA the uncertainties in $\mlhiggs$
and $m_t$ leave open the possibility that the total BR of the 4-body
decay of the $\lstop$-NLSP is sizable although the parameter space
corresponding to this scenario is severely squeezed by the bound 
on $\mlhiggs$. Perhaps the main future interest in this decay channel 
in the context of mSUGRA would be to observe it as a rare decay mode at 
the LHC or the ILC. Moreover, the existing bounds on $\mlstop$ obtained 
by assuming 100\% BR of the loop decay are approximately valid in most 
of the parameter space.

\begin{figure}[htb]
  \begin{center}
   \begin{tabular}{cc}
\hspace*{-3.3cm}   \mbox{\epsfxsize=0.8\textwidth
      \epsffile{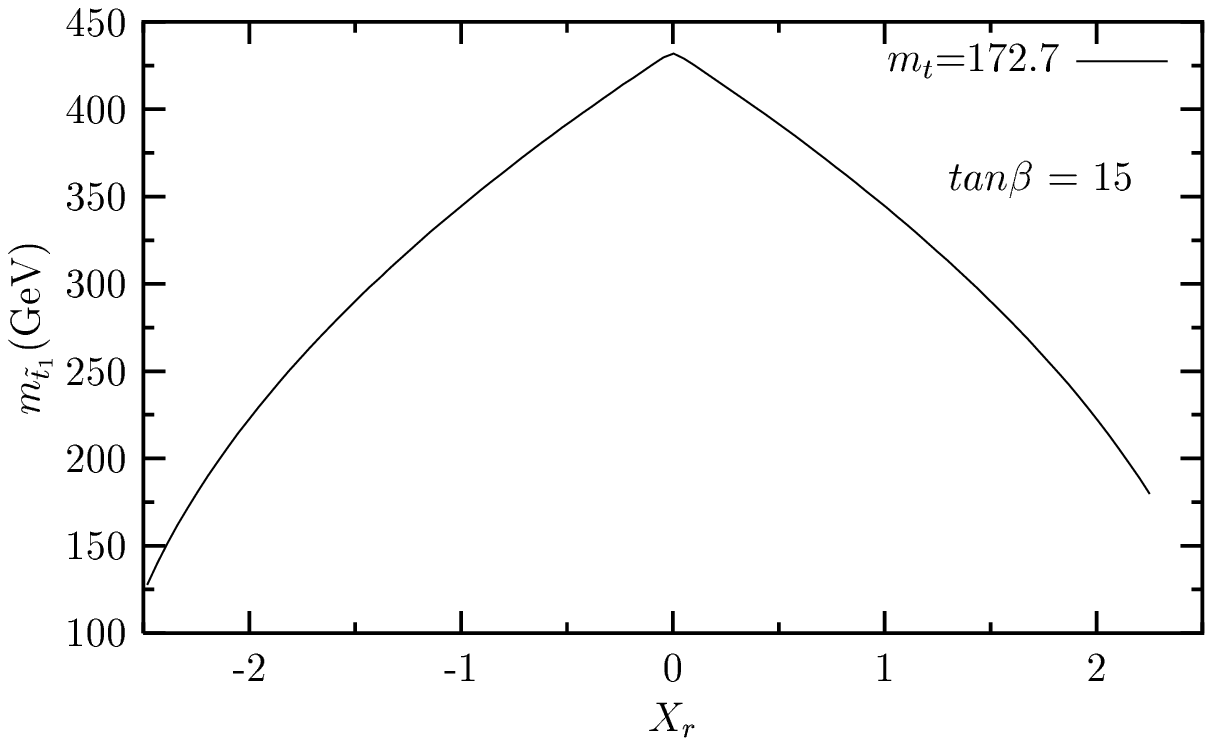}}
   &
\hspace*{-5.1cm}  \mbox{\epsfxsize=0.8\textwidth
      \epsffile{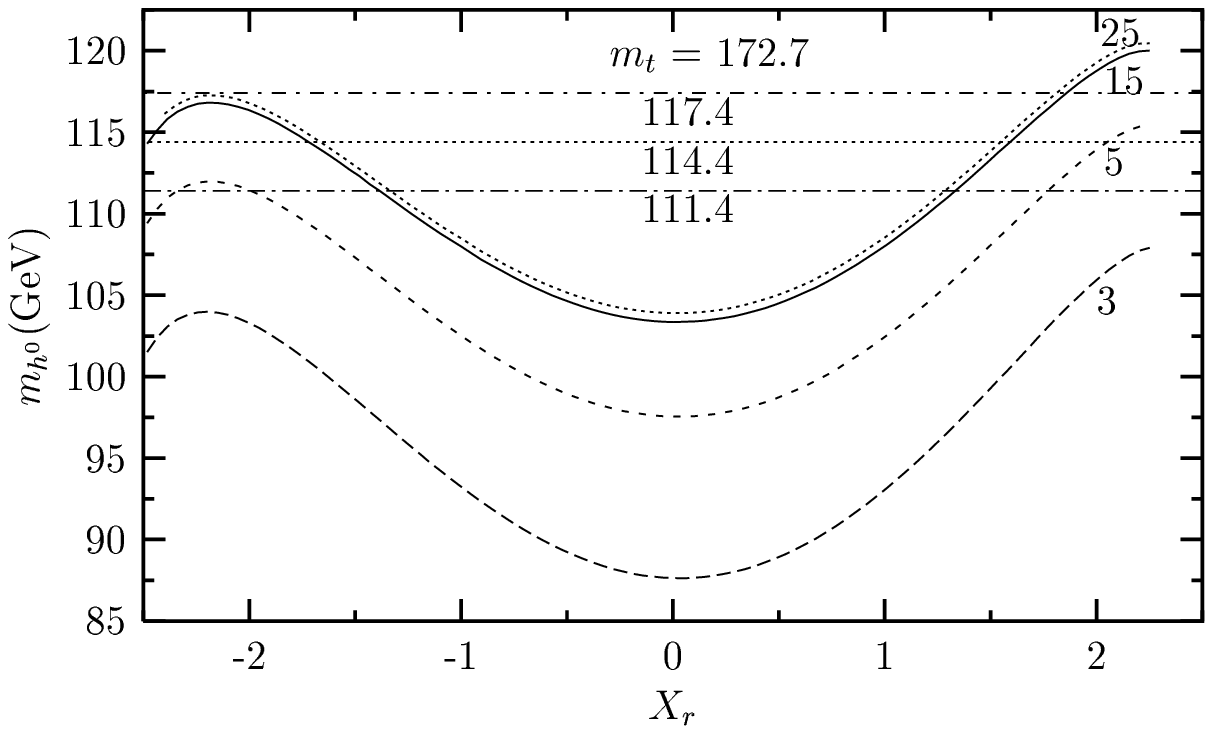}} 
  \end{tabular}
  \end{center}
\vspace{-11.4cm}
\caption{ \label{fig:xrtbmstmh}
{\small
The $\mlstop$($\mlhiggs$) as a function of $\xr$ in the 
MSSM in the left (right) panel for $m_t$=172.7. The MSSM model parameters  
are: $M_2$=300, $\mu$=300, $M_A$=500, $A_b$=200, $A_{\tau}$=300, 
$\msq=400$ and $\msell=250$. The choices of $\tb$ are shown in the  
legends. The $\lstop$-NLSP criterion is relaxed 
in this particular figure. 
}}
\end{figure}

\begin{figure}[htb]
  \begin{center}
   \begin{tabular}{cc}
\hspace*{-3.3cm}   \mbox{\epsfxsize=0.8\textwidth
      \epsffile{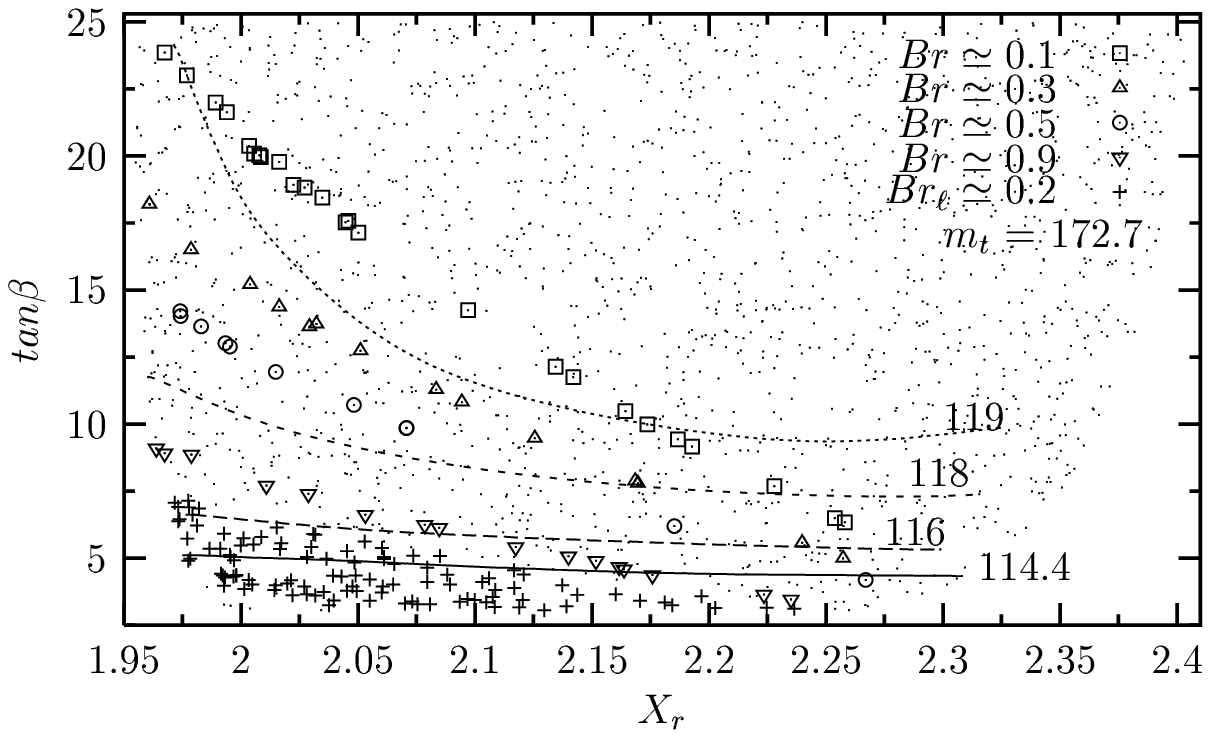}}
   &
\hspace*{-5.1cm}  \mbox{\epsfxsize=0.8\textwidth
      \epsffile{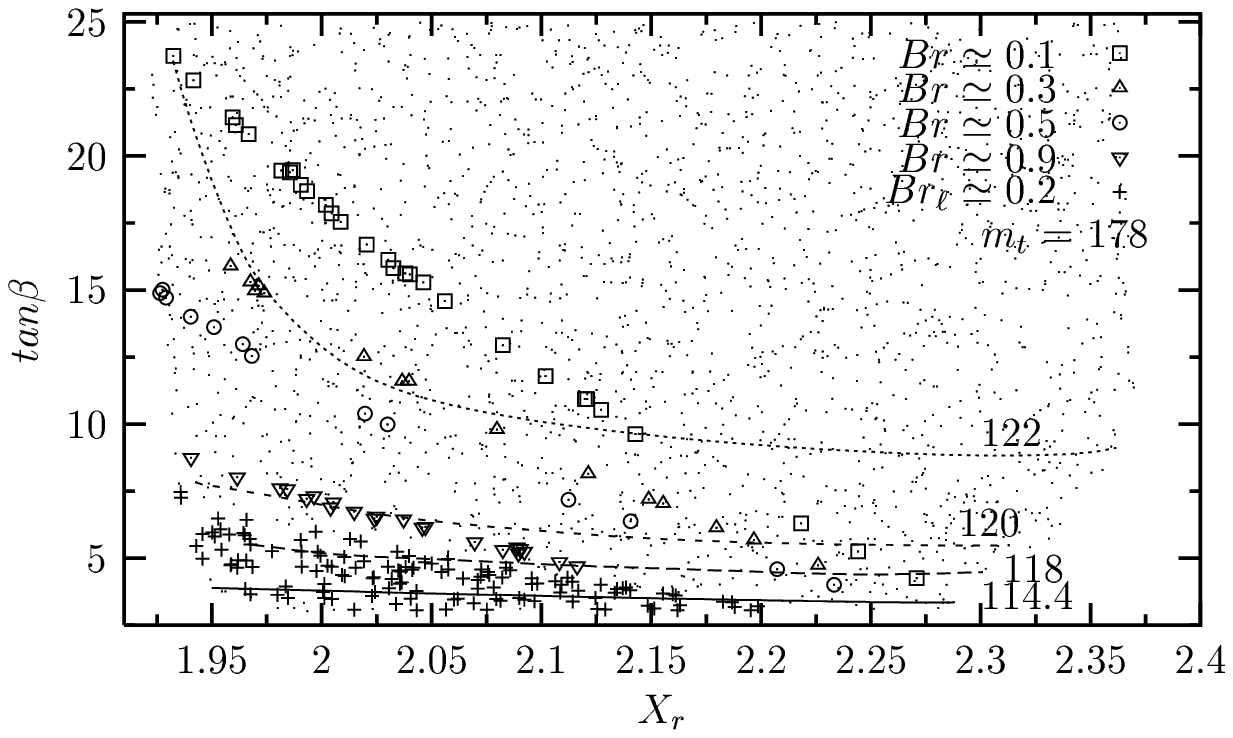}} 
  \end{tabular}
  \end{center}
\vspace{-11.4cm}
\caption{ \label{fig:xrtanbmtwo300msl250}
{\small
The $\lstop$ is the NLSP in the whole marked region  consistent 
with other sparticle mass bounds 
in the $\xr -\tb$ plane in the MSSM model for 
$M_2$=300, $\mu$=300, $M_A$=500, $A_b$=200, $A_{\tau}$=300, 
$\msq=400$, $\msell=250$ and $m_t$ =172.7(178.0) in the  left (right) 
panel. 
The total 4-body decay BRs of the lighter top-squark 
are shown by the contours with
different point styles. The `{\bf \small +}' shows 
the regions where $\brfourdklep \simeq 20\%$ for $\ell=e$ and $\mu$. The  
Higgs boson mass contours for 114.4, 116.0, 118.0 and 119.0 (114.4, 118.0, 
120.0 and 122.0 ) are also  plotted in the left (right) panel.  
}}
\end{figure}

\begin{figure}[htb]
  \begin{center}
   \begin{tabular}{cc}
\hspace*{-3.3cm}   \mbox{\epsfxsize=0.8\textwidth
      \epsffile{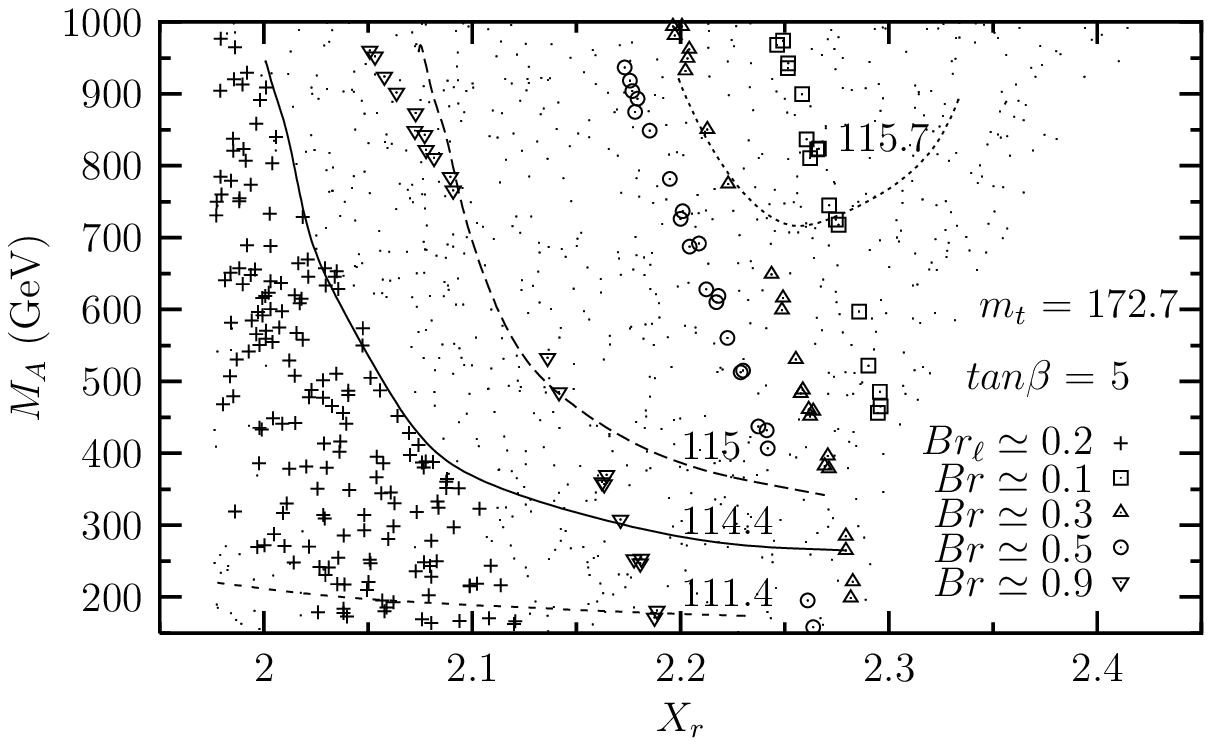}}
   &
\hspace*{-5.1cm}  \mbox{\epsfxsize=0.8\textwidth
      \epsffile{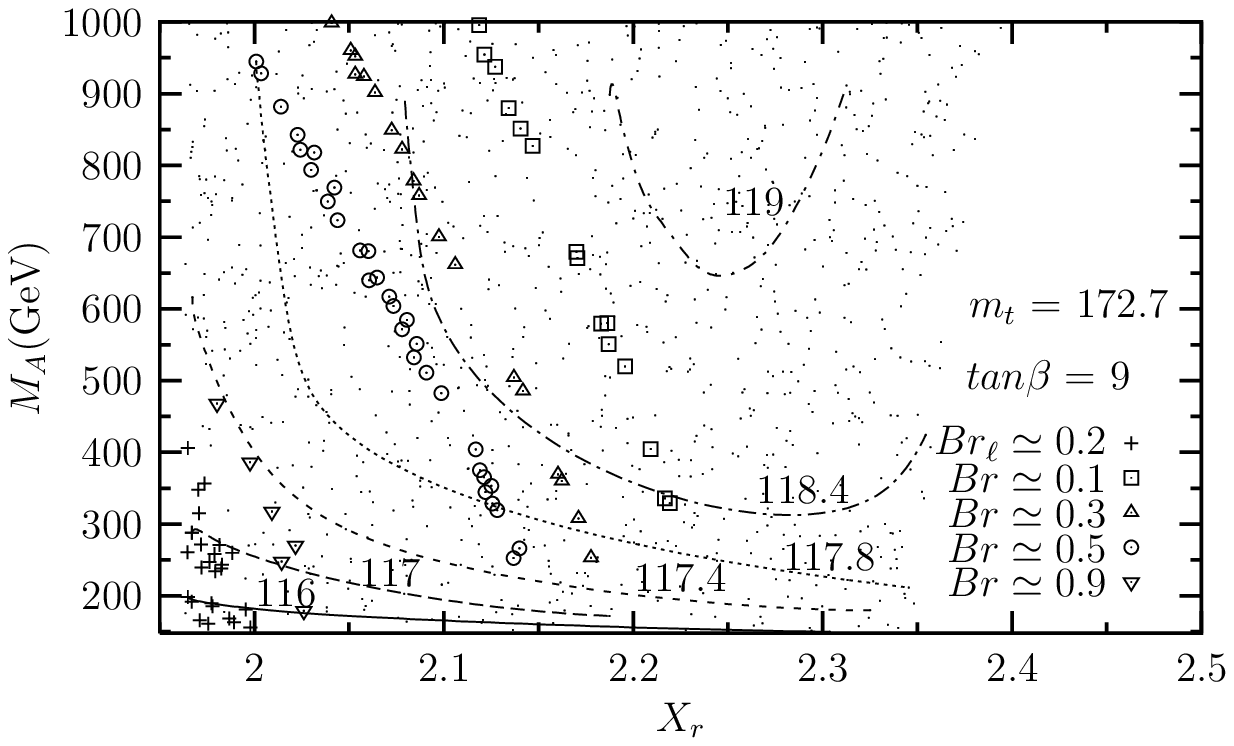}} 
  \end{tabular}
  \end{center}
\vspace{-11.4cm}
\caption{ \label{fig:xrma178tb59mup}
{\small
The $\lstop$ is the NLSP in the whole marked  region 
in the $\xr - M_{A}$ plane in the MSSM model for 
$M_2$ =300, $\mu$= 300, $m_{\wt{q}}$=400, 
$m_{\wt{\ell}}$=250, $A_{b}$=200, $A_{\tau}$=300, $m_t$ =172.7 and 
$\tb$=5(9) in the left (right) panel. The 
total 4-body decay BR of lighter top-squark 
are shown by different point styles. The `{\bf \small +}' shows 
the regions where $\brfourdklep \simeq 20\%$ for $\ell$=e and $\mu$. The Higgs boson  
mass contours for 111.4, 114.4, 115.0 and 115.7 (116.0, 117.0, 117.4, 117.8, 
118.4 and 119.0 ) are also plotted in the left (right) panel. 
}}
\end{figure}

In MSSM we have used the parameter $\xr$(see Eqn.\ref{xrmssm}) as an 
input. The interplay of the $\lstop$-NLSP condition and the Higgs boson 
mass constraints can be easily followed by varying  $\xr$ (see 
Fig.\ref{fig:xrtbmstmh}). One of the important conclusions is that sizable 
total BR of the 4-body decays of $\lstop$ can occur for the stronger constraints
$\mtop=172.7$, $\mlhiggs \gsim 114.4$ and intermediate values of 
$\tb$ (10 - 15) (see Fig.\ref{fig:xrtanbmtwo300msl250}). For example, this  
BR can be as large as 90\%  in the 
$\xr - \tb$ plane even for $\tb\approx 9$, $\mlhiggs \gsim 114.4$ and 
$\mtop=$ 172.7 and 178.0. For lower values of $\tb$
the leptonic 4-body BR $\simeq$ 20\%, which can lead to an observable
signal at Tevatron Run-II \cite{me1}, is also consistent with the 
above constraints. The total BR can be restricted to the level of
10 - 20 \% only if the bound on $\mlhiggs$ is significantly improved and $m_t$
is determined with better precision. Complementary information leading to
similar conclusions are presented also in the $\xr - M_{A}$ plane (see 
Fig.\ref{fig:xrma178tb59mup}).

In conclusion we emphasize  that the bounds on $\mlstop$ from Tevatron 
obtained  by assuming the complete dominance of 
the loop decay of $\lstop$ requires revision if we want these bounds to
be model independent. This is true even 
though the strong constraint on $\mlhiggs$ from LEP squeezes the 
parameter space which favours the competing 4-body decays of $\lstop$. 
Based on our earlier analyses we conclude that the  4-body mode could still 
be the main discovery channel for $\lstop$ at both Tevatron Run-II and LHC.  
As future strategies for the discovering $\lstop$ are planned, attention 
should be paid to the lighter CP-even Higgs boson search. If its mass 
bound is significantly improved, the interest in the 4-body decay would 
be to look for it as a rare decay. It is expected that the 
Higgs boson may show up at an early stage of LHC the experiments and,  
at least, some preliminary information on its mass will be available. This 
information will finally decide the viability of $\lstop$ search 
in the 4-body decay channel.

\noindent

{\bf Acknowledgement}: The author thanks Amitava Datta  and 
Amitava Raychaudhuri for suggestions, comments and careful 
reading of the manuscript. The author is 
thankful to M. Guchait, M. Maity and S. Poddar for discussions. A part of 
the work was supported by the project (SP/S2/K-10/2001) of the 
Department of Science and Technology (DST), India.

\newpage
\vspace{-3.0cm}
\vspace{-1.2cm}

\vspace{-3.0cm}
\vspace{-1.2cm}

\vspace{-3.0cm}
\vspace{-1.2cm}

\vspace{-3.0cm}

\vspace{-3.0cm}

\vspace{-3.0cm}

\vspace{-3.0cm}
\end{document}